\documentclass{PoS}

\title{Neutrino mass hierarchy and matter effects}

\ShortTitle{Neutrino mass hierarchy and matter effects}

\author{\speaker{Alexei Yu. Smirnov} \\        

        International Center for Theoretical Physics, Trieste, Italy\\
        E-mail: \email{smirnov@ictp.it}}

\abstract{Matter effects modify the mixing and the effective masses of neutrinos 
in a way which depends on the neutrino mass hierarchy. 
Consequently,  for normal and inverted hierarchies the  
oscillations and flavor conversion results  are different. 
Sensitivity to the mass hierarchy appears whenever the
matter effects on the 1-3 mixing and mass splitting become substantial.
This happens in supernovae in wide energy range 
and in the matter of the Earth. The Earth density profile is a multi-layer medium 
where the resonance and parametric enhancements 
of oscillations occur. The enhancement  is  realized  in neutrino 
(antineutrino) channels  for the normal (inverted) mass hierarchy.
Multi-megaton scale under ice (water) atmospheric
neutrino detectors with low energy threshold can establish mass
hierarchy with $(3 -  10) \sigma$  confidence level in few years.
The main challenges of these experiments 
are discussed and various ways to improve sensitivity are outlined.  
In particular, inelasticity  measurements will allow to
increase  significance of the hierarchy
identification by $20 - 50 \%$ .}

\FullConference{XV Workshop on Neutrino Telescopes,\\
		11-15 March 2013\\
		Venice, Italy}

\begin{document}

\def\be{\begin{equation}}
\def\ee{\end{equation}}
\def\bea{\begin{eqnarray}}
\def\eea{\end{eqnarray}}
\def\nnb{\nonumber}
\def\dgr{\dagger}

\section{Introduction}

The neutrino mass states can be marked by amount  
of the electron flavor  
in such a way that the $\nu_e$ admixture  
decreases with increase of the state number. 
The normal mass hierarchy (NH) corresponds then to the  ordering 1-2-3,  
whereas the inverted hierarchy (IH) has  
the ordering 3-1-2. Therefore  ``inversion'' of the hierarchy 
means actually the cyclic permutation of the mass states. 
The hierarchy is related to the sign 
of $\Delta m_{31}^2 \equiv m_3^2 - m_1^2$. 

Establishing the mass hierarchy is one of the key objectives 
in neutrino physics. The mass hierarchy has important  phenomenological consequences  for the  
supernova neutrinos, for cosmology, for the atmospheric 
and accelerator neutrinos propagating in the matter of the Earth. 
The hierarchy is important whenever the matter 
effects  on  oscillations or conversion 
driven by the 1-3 mixing and mass splittings are important \cite{blennow}. 
The strongest effect is in the resonance region 
where  
\be
E \approx \frac{\Delta m_{31}^2}{2 V},  
\label{xparam}
\ee
and  $V$ is the matter potential. In particular, for the Earth matter density the relevant interval 
is $E \sim (4 - 10)$ GeV.    

From theoretical point of view the NH case is closer to the situation 
in the quark sector with certain rescaling of mixing. 
NH can be easily obtained from the seesaw and with NH realization of the 
quark - lepton symmetry and unification is simpler. 
In contrast, the IH implies strong degeneracy of the two mass states with $\Delta m_{21}/ m_1  \leq \Delta m_{21}^2/ (2\Delta m_{31}^3) \leq 1.6 \cdot 10^{-2}$. 
This, in turn, indicates certain flavor symmetry. 
In the first approximation the mass spectrum can be considered 
as a combination of one pseudo-Dirac neutrino and one Majorana neutrino. The former implies maximal 1-2 mixing 
and substantial deviation from maximal one follows from the charged 
leptons. 

The race for the mass hierarchy has been started.  
The idea is to explore the matter effects on the 1-3 mixing. 
This can be realized  by studying propagation of 
high energy neutrinos inside the Earth (see, e.g. \cite{hiermatt}) . 
The possibilities include  (i) Detection of   atmospheric neutrino 
fluxes with magnetized spectrometers like INO  
\cite{INO} 
or with huge atmospheric neutrino detectors 
PINGU~\cite{PINGU},  ORCA~\cite{ORCA},  HyperKamiokande~\cite{HK};  
(ii) Long base-line (LBL) experiments 
NO$\nu$A~\cite{NOVA}, LBNE~\cite{LBNE}, LBNO~\cite{LBNO}.    
The  ultimate proposal here could be the LBL experiment Fermilab -  PINGU~ \cite{winter} 
for which neutrinos will cross the core of 
the Earth and undergo the parametric enhancement of oscillations; 
(iii) detection of low energy supernova neutrinos.   

The paper is organized as follows. 
We first (sec. \ref{sec2}) consider mixing and level splitting 
in matter and their dependence on the mass hierarchy. 
Signatures of two hierarchies in supernova neutrinos 
will be summarized in sec.~\ref{sn}.
Sec. \ref{sec4} is devoted to effects of neutrino 
propagation in the matter of the Earth. 
A possibility to establish the hierarchy with huge atmospheric neutrino detectors
will de discussed in sec. \ref{hand}. Different ways to improve sensitivity 
of these detectors to the hierarchy will be presented in sec. \ref{impr}. 
Sec. \ref{conc} summarizes the results.  

\section{Mixing and level splitting in matter}
\label{sec2}

The mixing is determined with respect to the flavor states 
$\nu_f =   (\nu_e, \nu_\mu, \nu_\tau)$, and mixing 
matrix connects the flavor states with the  eigenstates 
of the Hamiltonian. Thus, the vacuum mixing connects 
the flavor  states with the mass states:
$\nu_f = U_{PMNS} \nu_{mass}$, whereas  
the mixing in matter connects the flavor states with 
the eigenstates of the Hamiltonian in matter: 
$\nu_f = U_{PMNS}^m \nu_H$.  

\begin{figure}
\includegraphics[width=2.9in]{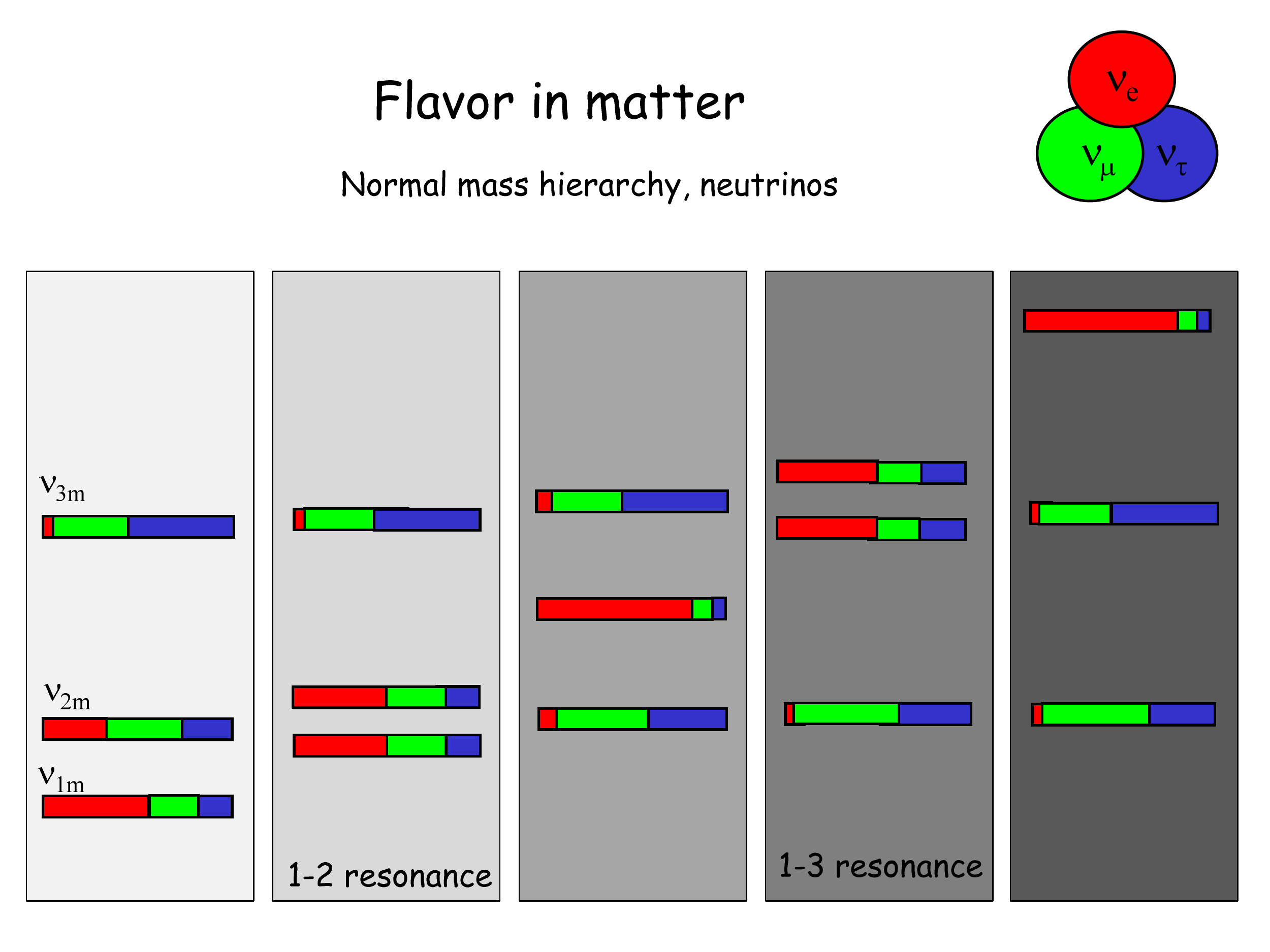}\hskip 3mm
\includegraphics[width=2.9in]{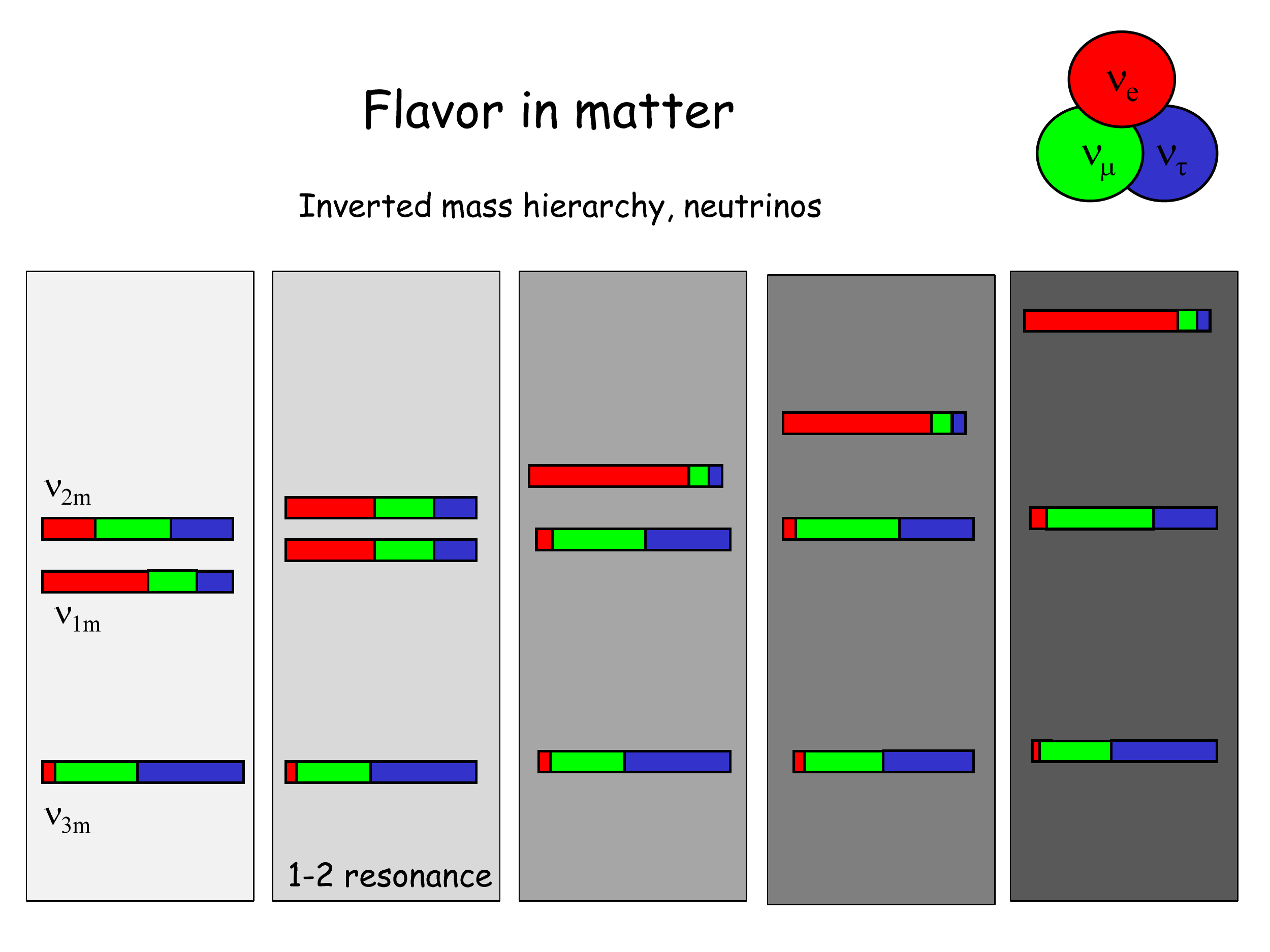}
\caption{Dependence of the neutrino mass and flavor 
spectrum on  matter density for NH ({\it left panel}), IN ({\it right panel}).
The density increases from  the left to right. }
\label{fff-1}
\end{figure}

\begin{figure}
\includegraphics[width=2.9in]{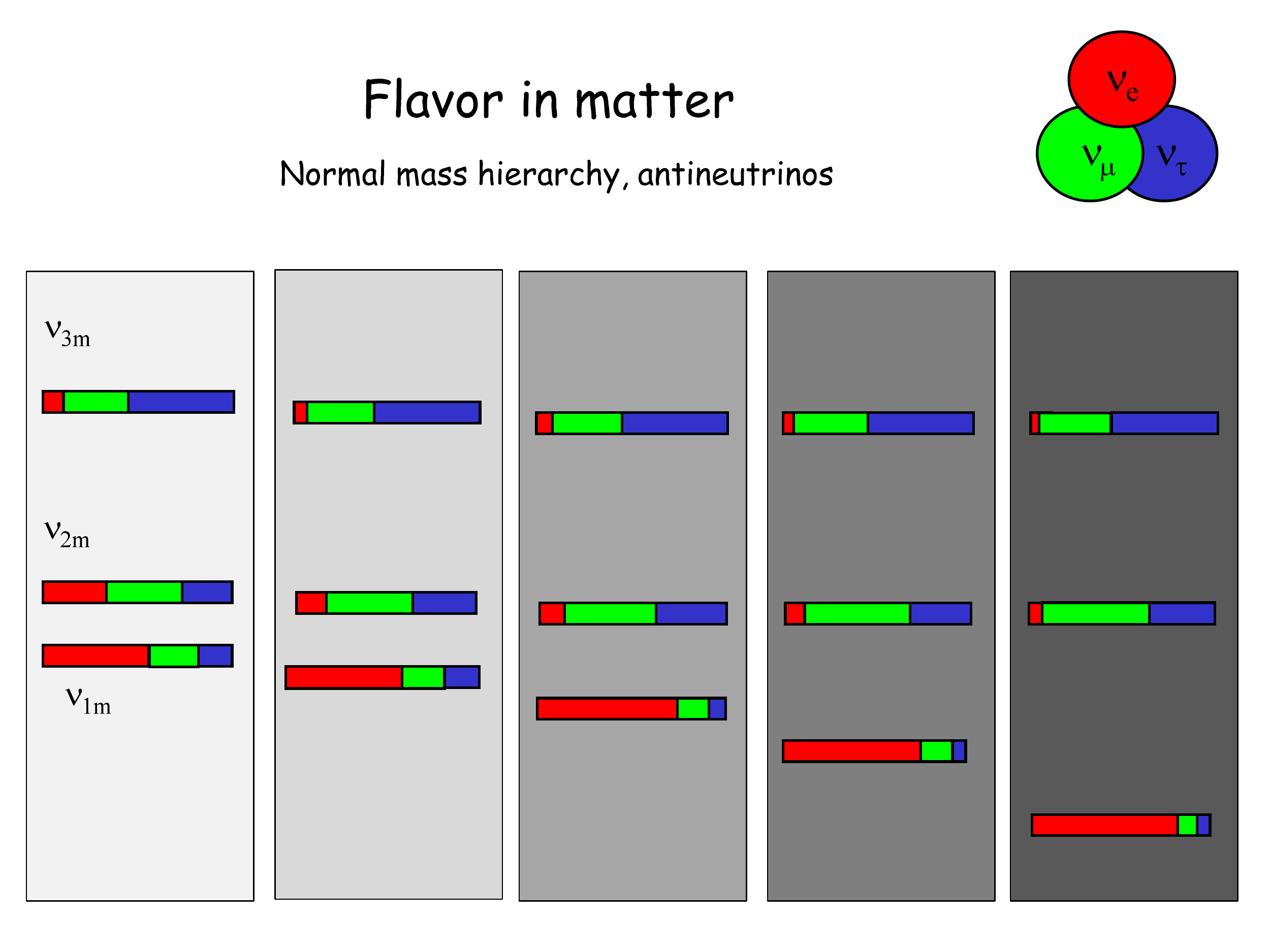}\hskip 3mm
\includegraphics[width=2.9in]{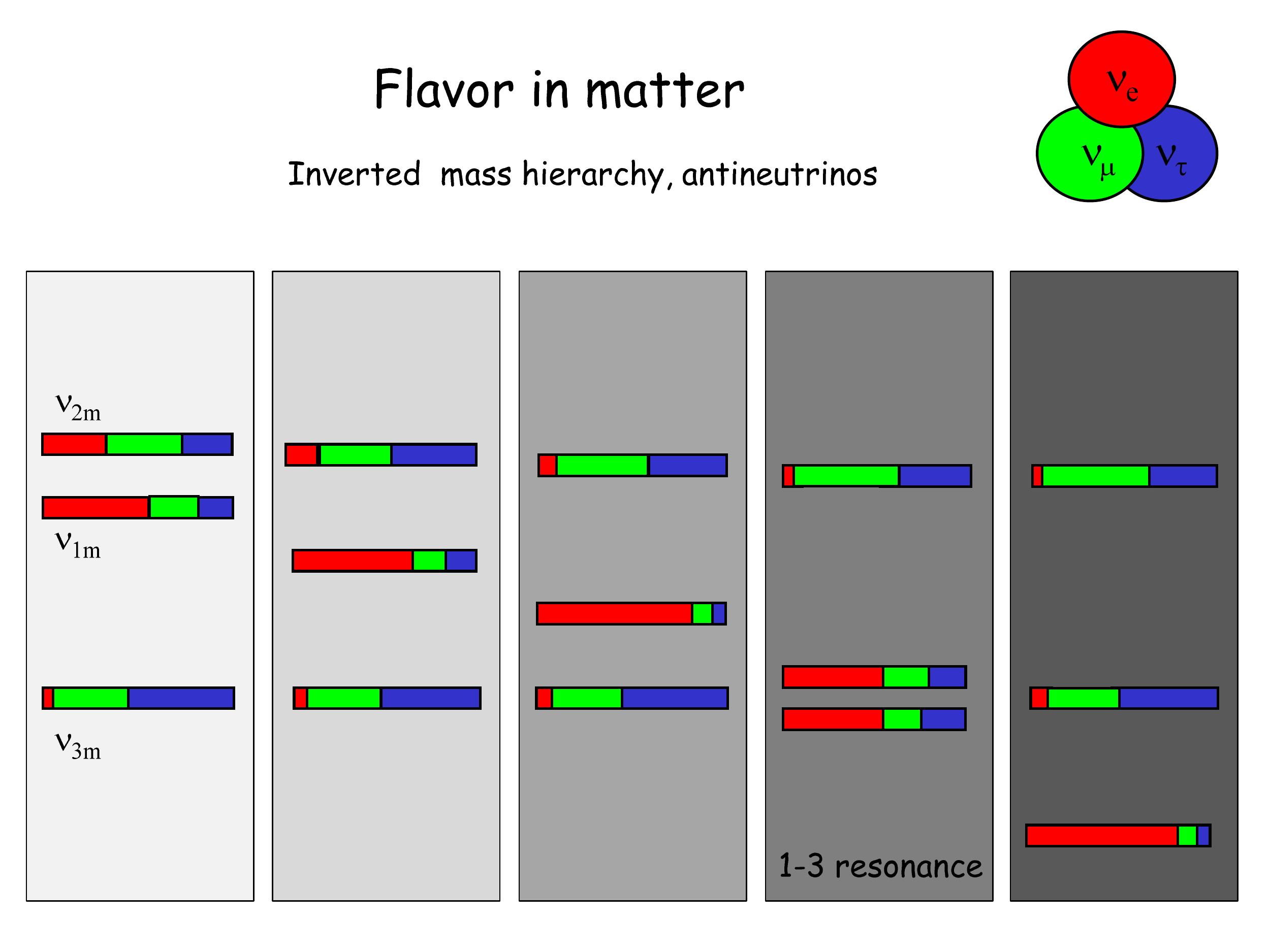} 
\caption{
The same as in Fig. 1 for antineutrinos. }
\label{fff-2}
\end{figure}

The mixing (flavor composition of the eigenstates) 
and the level splitting depend on density of matter and 
neutrino energy,  and these dependences 
are different for NH and IN.  
Furthermore, the  mass and flavor spectra change differently 
for neutrinos and antineutrinos.  In Fig.~\ref{fff-1} 
we show the neutrino  spectra for different densities 
in the cases of normal (left)  and inverted (right) hierarchies.  
For antineutrinos  the spectra are shown in Fig.~\ref{fff-2}.   
In vacuum spectra for neutrinos and antineutrinos differ 
for non-zero $\delta_{CP}$. Specifically, the distribution of the 
$\nu_\mu$ and $\nu_\tau$ flavors in the $\nu_1$ and $\nu_2$ 
are different. Resonances correspond to  configuration when 
two levels have small splitting and equal admixtures of the 
electron flavor. 

With increase of density the electron flavor 
becomes heavier, so that for NH it shifts from the lightest  state 
to the heaviest one. It  passes through two resonances,  
and at large densities $\nu_e$  concentrates in the heaviest state. 
For IH the change is less significant: $\nu_e$ ``passes'' 
through one resonance only.  
At very high densities the effective mass and flavor 
spectra are  similar for both hierarchies.  
They would be the same for maximal 2-3 mixing. 
The strongest difference is in the range of 1-3 resonance. 

In the antineutrino channel with increase of density 
the electron flavor becomes lighter. 
It shifts from the highest to the lowest energy level. 
There is no resonance in the case of NH, whereas 
for inverted hierarchy the 1-3 resonance is realized and the change is more substantial.  
Again the spectra for both hierarchies are similar at high densities. 

Inversion  of the  mass hierarchy is related 
to the neutrino - antineutrino interchange \cite{parke}.  
Indeed, the $\nu - \bar{\nu}$ substitution 
changes  sign of the potential: 
$V \rightarrow - V$. On the other hand, in the $2\nu-$ approximation 
the inversion of the mass hierarchy 
means  $\Delta m^2 \rightarrow  - \Delta m^2$. 
Therefore under simultaneous transformations  
$ NH \rightarrow  IH$ and 
$\nu \rightarrow \bar{\nu}$  the quantity 
$\Delta m_{31}^2/(2E V)$, 
which determines characteristics of oscillations in matter,  
is invariant.  This means that both mixing  and   
moduli of the oscillation 
phase do not change. Thus, in the $2\nu$ case  
\be
P^{NH} = \bar{P}^{IH}, ~~~~P^{IH} = \bar{P}^{NH}.
\ee
Consequently,  the difference of numbers of events for NH and IH,  
$\Delta N  \equiv  N^{NH} - N^{IH} = \sigma F (P^{NH} - P^{IH})$,  
has opposite sign for neutrinos and antineutrinos: 
\be
\Delta \bar{N}  \equiv  \bar{N}^{NH} - \bar{N}^{IH} 
= \bar{\sigma}\bar{F} (\bar{P}^{NH} - \bar{P}^{IH}) = 
- \bar{\sigma}\bar{F} (P^{NH} - P^{IH}) = 
- \frac{\bar{\sigma}\bar{F}}{\sigma F} \Delta N ,  
\ee
where bars indicate  characteristics for antineutrinos. 
Therefore when summed the  neutrino and antineutrino  
signals cancel each other  partially   
because of difference of cross-sections and fluxes 
for neutrinos and antineutrinos.  

Let us consider effect of hierarchy on solar neutrinos. Recall that 
the 1-2 ordering is normal and this has been established 
due to matter effects on the 1-2 mass splitting and mixing using 
the solar neutrinos. Essentially the hierarchy is fixed by the fact 
that suppression of the neutrino flux in the high energy part of the spectrum 
is weaker than in the  low energy part. For the survival probability we have   
for low energies $P_{ee} \approx 0.5 \sin^2 2 \theta_{12} \approx 0.58$ and for high 
energies:  
\be
P_{ee}  = |U_{e2}|^2 \approx \sin^2 \theta_{12} \approx 0.3~~~ 
(NH), ~~~~~ 
P_{ee}  =  |U_{e1}|^2 \approx \cos^2 \theta_{12} \approx 0.66~~~(IH).  \ee

The solar neutrinos have low  sensitivity to the 1-3 ordering 
because of  small matter effect on the 1-3 mixing:   
$
\sin \theta_{13}^m \approx \sin \theta_{13} 
(1 \pm (2EV/\Delta m_{31}^2));  
$
here $+/-$ correspond to NH/IH. Then the correction to the 
survival probability due to matter effect on the 1-3 mixing equals~\cite{solar} 
\be
\frac{\Delta P_{ee}}{P_{ee}} \approx \mp 2 
\sin^2 \theta_{13} \frac{2EV_c}{\Delta m_{31}^2},   
\label{h-solar}
\ee
where $V_c$ is the value of potential in the neutrino production 
region in the Sun. Minus sign corresponds to NH and 
the correction increases with energy.   
For $E = 10$ MeV we obtain from  (\ref{h-solar}) 
$\Delta P_{ee} \sim 10^{-3}$ 
which  is below the sensitivity of existing experiments on solar neutrinos.

\section{Supernova neutrinos}
\label{sn}

For supernova neutrinos the matter effects dominate,  
and consequently,  the type of hierarchy is crucial. 
With measured value of the 1-3 mixing 
the level crossing in the H- (high density) resonance 
is highly adiabatic. This removes many 
ambiguities and the picture of flavor 
conversion in the MSW resonance region becomes very simple. 
Adiabaticity can be broken in shock wavefront only. 
Results of the  MSW conversion  can be affected also by the collective 
neutrino effects that happen in the deeper regions of a star.

The following hierarchy-sensitive effects can be observed.

1) Shock wave breaks adiabaticity of the flavor conversion in the 1-3 resonance.  
This leads to  softening of the spectrum of the electron neutrinos  
since $\nu_e \rightarrow \nu_{\mu, \tau}$ conversion becomes less efficient in certain energy interval. 
The interval shifts with time (during the burst) from low to high energies~\cite{tomas}.  
Observation of this effects in the neutrino (antineutrino) 
channel will imply normal (inverted) hierarchy.   

2) Neutrino collective effects are   
more profound in the IH case and can  be realized 
when neutrino density near the core of a star becomes 
comparable or larger than usual density. 
If the spectral splits are observed at high energies, 
the hierarchy  should be inverted~\cite{fuller},~\cite{dasgupta}.

3) Sharp time-rise of the  $\bar{\nu}_e$ flux and signal    
in the initial phase of neutrino burst will testify for IH  
\cite{serpico}. 

4) Strong suppression of the $\nu_e-$ neutronization peak 
is the signature of  NH.  In this case  $\nu_e \rightarrow \nu_3$ 
transition occurs and the $\nu_e$ survival probability 
equals $P_{ee} = \sin^2 \theta_{13} \approx 
0.02$, as compared to  $P = \cos^2 \theta_{12} \approx 0.68$ 
in the case of IH~\cite{dighe}.

5) At the accretion and cooling phases a strength of partial permutation of the electron and 
non-electron neutrino spectra depends on the type of mass hierarchy. 
As a result, the $\nu_e$ energy spectrum (and similarly 
the $\bar{\nu}_e$ spectrum) 
becomes two-component: 
a mixture of the original  $\nu_e$ and $\nu_\mu$ spectra. 
Precise composition depends on the  mass hierarchy~\cite{dighe}, \cite{lunardini}. 

6) The Earth matter effects consist of  
an oscillatory modulation of the neutrino energy spectrum and difference of signals in 
detectors situated in different places of the Earth \cite{dighe}, \cite{earth}.  
These effects are due to the 1-2 mixing, 
however their appearance depends on conversion in a star  
driven by the 1-3 mixing.  
Being observed in the antineutrino channel 
the effects  will be the evidence of NH,  
if this happens  in the neutrino channel,  IH  
is established. 

The problem here is that in the antineutrino channel,  
which is the most suitable for detection, the difference 
of original fluxes of the electron and non-electron 
antineutrinos, and consequently, the oscillation effects 
are small.

\section{Propagation in the Earth}
\label{sec4}

\begin{figure}
\begin{center}
\includegraphics[width=4.7in]{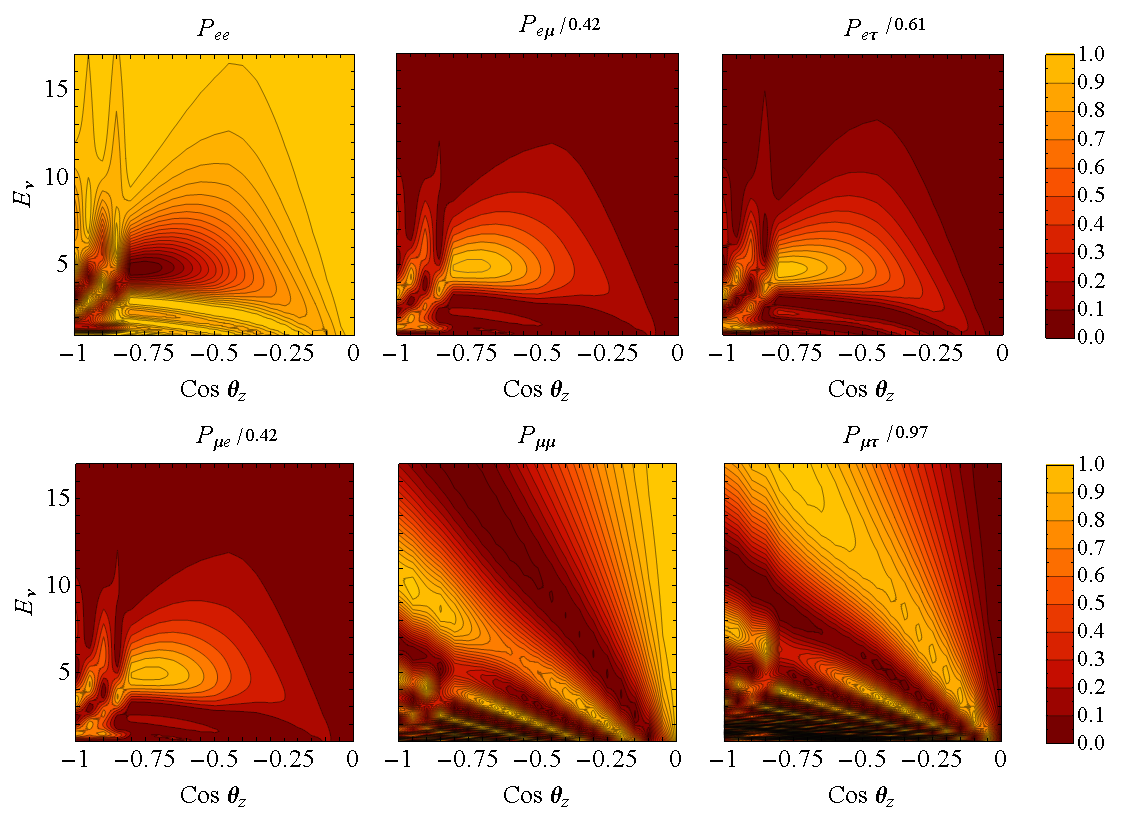}
\end{center}
\caption{
Neutrino oscillograms of the Earth for different oscillation
channels for normal mass hierarchy. Shown are the oscillation
probabilities normalized by their maximal values (from \cite{ARS}).}
\label{fff-3}
\end{figure}

Inside the Earth we deal with oscillations in multi-layer medium. 
The oscillation effects are best presented in the oscillograms -- 
lines of equal oscillation probabilities in the 
($\cos \theta_z - E$) plane, where  $\theta_z$   is the zenith angle of neutrino trajectory 
\cite{osc1}, \cite{maltoni} (see Fig.~\ref{fff-3} from \cite{ARS}). 
The oscillograms depend on mass hierarchy and inverting 
the hierarchy means approximately an interchange of the $\nu$ and  $\bar{\nu}$ 
channel, especially at high energies. Thus, propagation of neutrinos with $E \sim (3 - 20)$ GeV  
can be used to establish the hierarchy. 
According to Fig.~\ref{fff-3} constructed for neutrinos,  
salient features of the oscillograms are 
\begin{itemize}

\item
the resonance peak (for $P_{e\mu}$, $P_{e\tau}$, $P_{\mu e}$) 
or dip (for $P_{ee}$) in the mantle domain ($\cos \theta_z > - 0.83$) at $E \sim 6$ GeV and $\cos \theta_z = - 0.8$; 

\item  

three parametric ridges in the core  domain 
($\cos \theta_z < - 0.83$) at $E_\nu = (2 - 10)$ GeV.   

\end{itemize}

\begin{figure}
\includegraphics[width=3.0in]{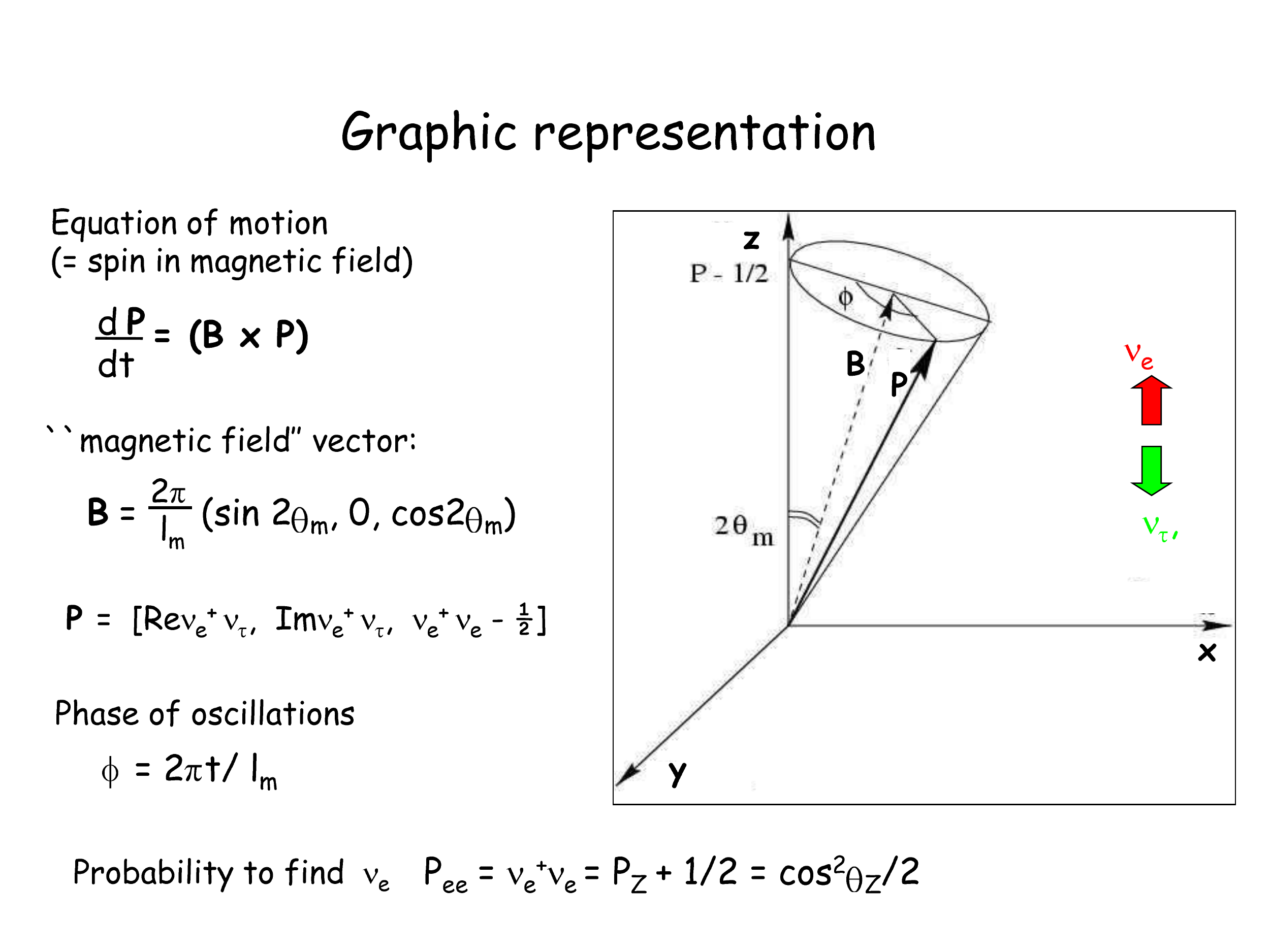}
\includegraphics[width=2.9in]{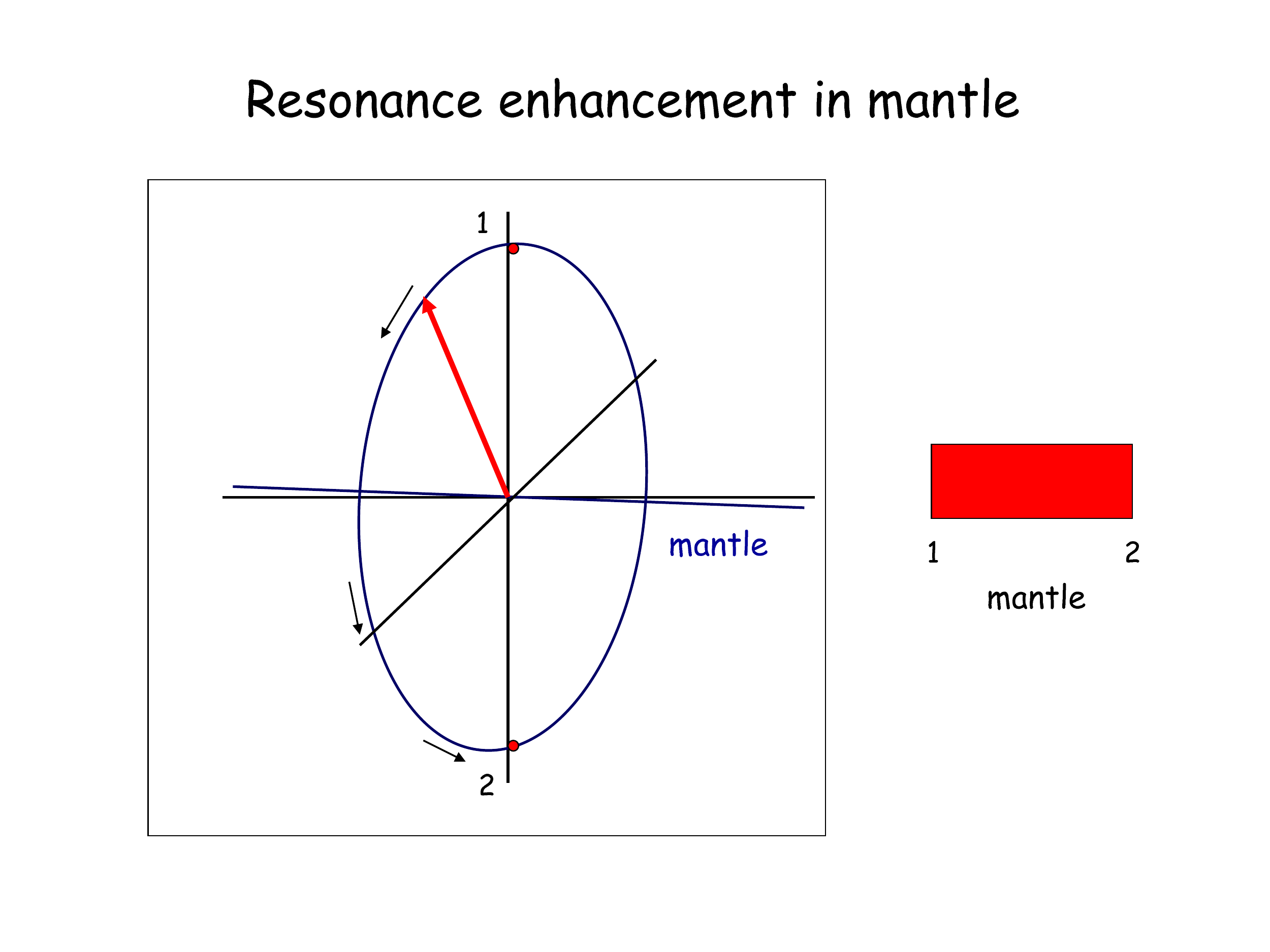}
\caption{
Graphic representation of the neutrino oscillations.
{\it Left panel:} generic case with explanations;
{\it right panel:} motion of the neutrino
vector which corresponds to the peak
in oscillogram due to the
resonance enhancement of  the $\nu_e - \nu_\tau^{\prime}$
oscillations in mantle. The oscillation phase
equals $\pi$.
}
\label{graph1}
\end{figure}

The most transparent and easiest way to understand these effects 
is to use graphic representation based on  analogy 
of the neutrino oscillations 
with  the  electron spin precession in the magnetic field (see Fig.~\ref{graph1}, left).  
(For definiteness we consider the two neutrino system,   
$\nu_e$ and  $\nu_{\tau}^{\prime}$, with $\nu_{\tau}^{\prime}$ being certain mixture of $\nu_\mu$ and $\nu_\tau$.) 
Precession of the neutrino vector ${\bf P}$ leads to periodic change of its projection 
onto axis $z$,  which is  equivalent to oscillations. 

\begin{figure}
\includegraphics[width=2.9in]{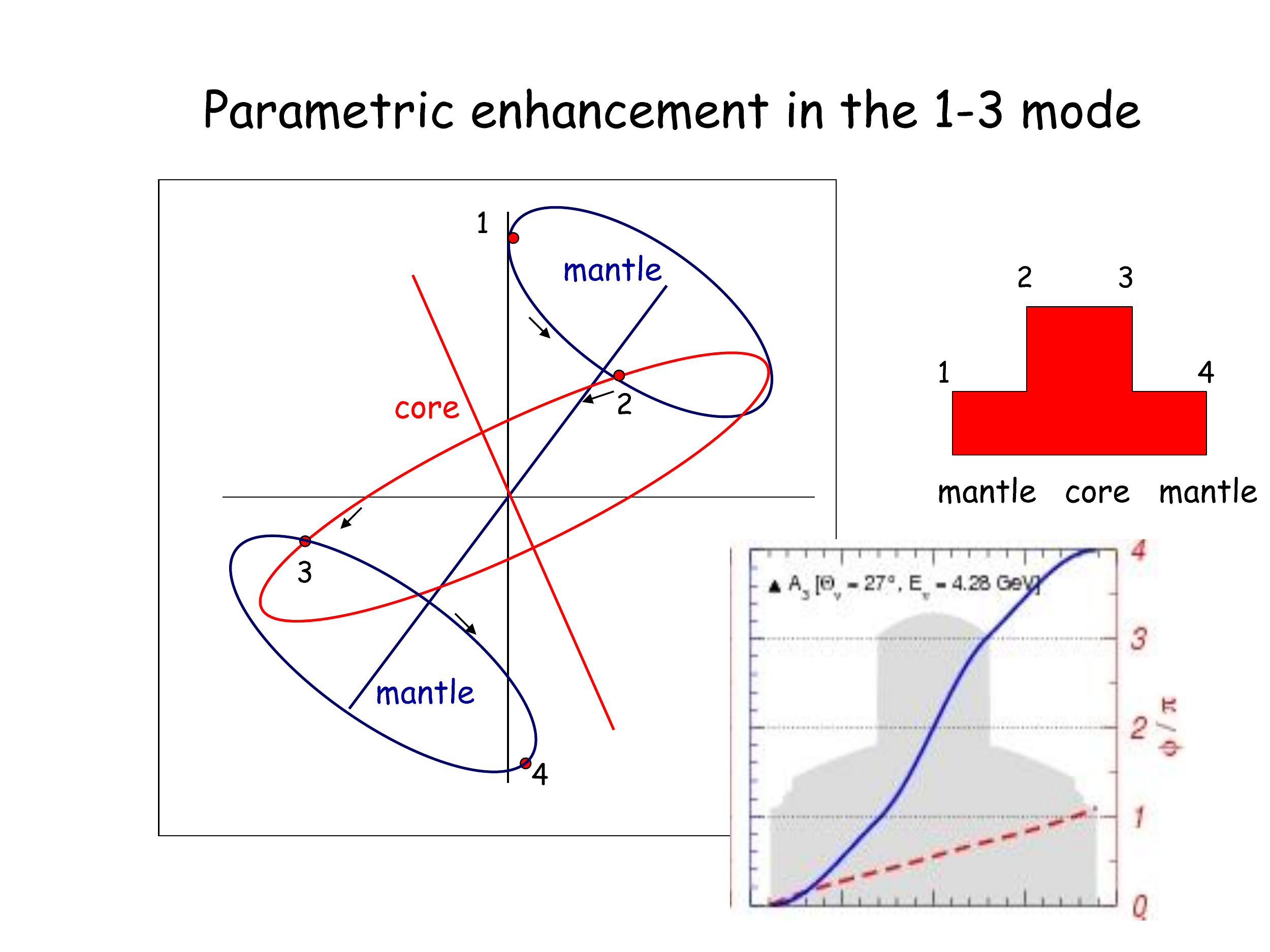}
\includegraphics[width=2.9in]{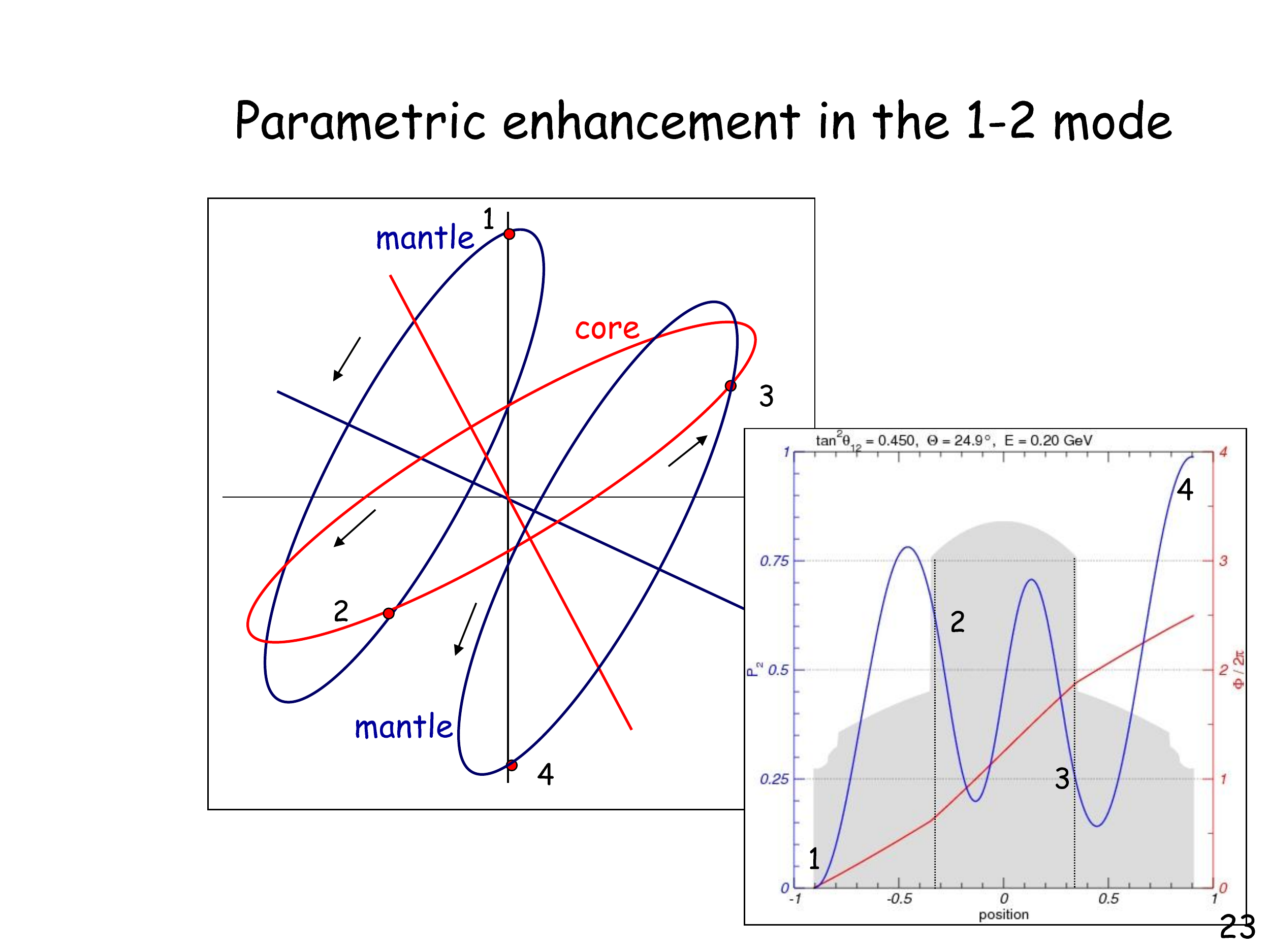}
\caption{
Graphic representation of the
parametric enhancement of the $\nu_e - \nu_\tau^{\prime}$ oscillations.
{\it Left panel:}  motion of the neutrino 
vector which corresponds to the parametric ridge in
oscillogram due to  the  1-3 mixing.
In the lower box shown  is the probability
as function of distance  (from \cite{maltoni}). 
{\it Right panel:} the same as in the left panel but for oscillations driven
by the 1-2 mixing.
}
\label{graph2}
\end{figure}
In Figs. \ref{graph1},  \ref{graph2} we show graphic representations of the main 
effects involved in propagation of neutrinos 
inside the Earth: 

1. Resonance enhancement of oscillations (Fig.~\ref{graph1},  right). 
The MSW resonance condition is  fulfilled in the mantle at 6 GeV. 
In resonance $2\theta_m = \pi/2$,  so that the cone axis is directed 
along the axis $x$ and precession occurs 
with maximal amplitude. For the zenith angle  $|\cos \theta_z | = 0.8$ 
the length of trajectory 
in the mantle is such that $L = l_m /2$ and therefore 
the oscillation phase equals $\pi$, so that in final state ${\bf P}_z = -1/2$. 
This corresponds to complete flip of ${\bf P}$    
and maximal probability of the $\nu_e \rightarrow \nu_\tau^{\prime}$ transition. 
and happens in  the peak in the oscillogram at $E = 6$ GeV.

2. Parametric enhancement of oscillations 
(Fig.~\ref{graph2}) occur for the core crossing trajectories when 
neutrinos experience propagation through three layers with slowly changing density and 
two density jumps on the border between the layers \cite{param}. This  leads to the parametric ridges. 
The cone axis  has different directions in the mantle and core (Fig.~\ref{graph2}). 
The axis changes its direction suddenly at the 
border between the mantle and the core. 
So,  the vector ${\bf P}$ precesses in the core and mantle around two different directions.  
For certain $E$ and $\theta_z$ the phases in the core and mantle 
can be such that the probability of transition builds up to maximal one  without return  and oscillations (see lower boxes). 
This produces strong transition even if in each layer 
the transition is  small.  Two such  possibilities are shown in Fig.~\ref{graph2}.   

\section{Hierarchy with Huge atmospheric neutrino detectors}
\label{hand}

The atmospheric neutrino fluxes, being cost free,  cover complete zenith angle range and 
huge interval of energies which includes the resonance region relevant 
for determination of the hierarchy.   
The original fluxes produced in atmosphere contain $\nu_\mu$ and $\nu_e$ 
and corresponding antineutrinos. Furthermore,  
the flavor content changes with the neutrino  energy.  
The main challenges here are (i) uncertainties of the original fluxes; 
(ii) reconstruction  of the neutrino  direction; 
(iii) the neutrino energy resolution; (iv) flavor identification. 
According to earlier 
estimations~\cite{mena} the  DeepCore detector with the energy threshold above 10 GeV 
has rather low sensitivity to the hierarchy.
High statistics and even mild technological developments
which will be achieved in  Multi-megaton under ice (water) cherenkov
detectors with relatively low energy threshold
$E_\nu \sim (2 - 3)$  GeV can resolve these problems~\cite{ARS}, \cite{winter-at}. 
Thus, PINGU detector \cite{PINGU} will have up to  
$10^5$ events in the  range (2 - 20) GeV which 
covers the 1-3 resonance region.

\begin{figure}
\begin{center}
\includegraphics[width=3.7in]{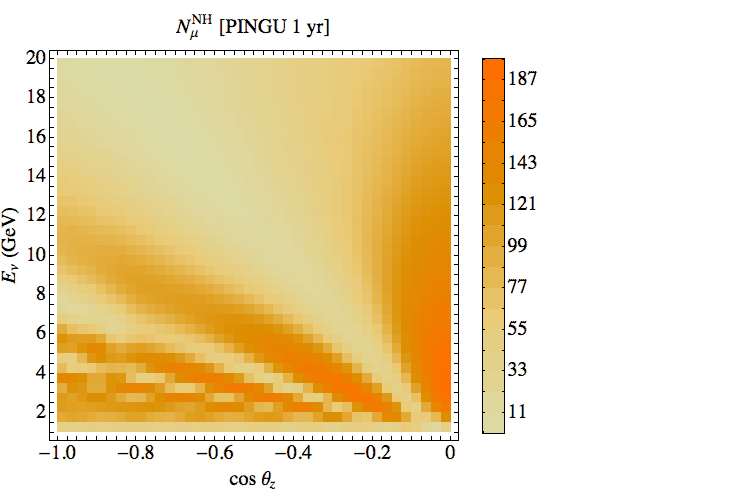} \hskip-2cm
\includegraphics[width=2.8in]{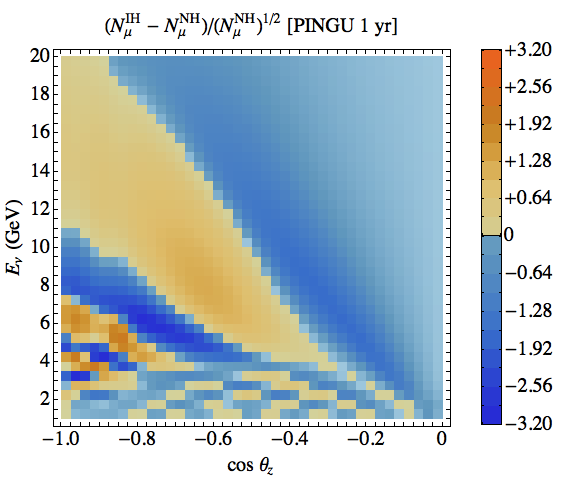}
\end{center}
\caption{
{\it Left panel:} The distribution of the  $\nu_\mu$ events  in the
($E_\nu - \cos \theta_z$) plane that can be collected
by PINGU detector during about 3 years. NH is  assumed.
{\it Right panel:} The hierarchy asymmetry of $\nu_\mu$ events in the
$E_\nu - \cos \theta_z$ plane without smearing (adapted from \cite{ARS}).}
\label{dist}
\end{figure}

Important sample of events is due to the $\nu_\mu-$ charged current interactions 
(the $\mu-$ track events): 
$\nu_\mu + n \rightarrow \mu + h$. With  
densely instrumented detector it is possible to measure 
the muon energy $E_\mu$ and muon direction, $\theta_\mu$, $\phi_\mu$, 
as well as the energy of hadron cascade, $E_h$.  
This becomes possible analyzing  
time development of the event.  
Consequently, the neutrino energy, $E_\nu = E_\mu + E_h$ and, to some extend,  
the neutrino direction can be reconstructed. In Fig.~\ref{dist} (left)
we show a distribution of the $\nu_\mu$ events 
in the $E_\nu - \cos \theta_z$ plane.

Quick estimation of discovery potential can be obtained 
using  the hierarchy (H-) asymmetry. For each $ij-$bin
in the ($E_\nu - \cos \theta_z$) plane the H-asymmetry  is defined as \cite{ARS} 
\be
S_{ij} = \frac{N^{IH}_{ij} - N^{NH}_{ij}}{\sqrt{N^{NH}_{ij}}}.  
\label{disting}
\ee
If NH is the true hierarchy,  $N_{ij}^{NH}$     
can be considered as the ``experimental'' number of events,    
whereas  $N_{ij}^{IH}$ -- as the ``fit''  number  of events.
Then  $|S_{ij}|$ reflects statistical 
significance of establishing true hierarchy. 
Clearly this quantity does not take 
into account fluctuations and therefore 
more appropriate term could be {\it distinguishability}. 
Still $S_{ij}$  is very useful characteristic which allows one  
to study dependence of the discovery 
potential on values of involved 
parameters, uncertainties, degeneracies, {\it etc.}. 

The uncorrelated systematic errors can be introduced 
adding to the denominator of (\ref{disting}) 
\be 
N_{ij}^{NH} \rightarrow \sigma_{ij}^2 =
N_{ij}^{NH} + (f N_{ij}^{NH})^2 , 
\label{syst}
\ee 
where $f$ determines the level of systematic errors. 
If measurements in each bin 
are independent the total significance is then given by 
\be
S_{tot} = \sqrt{\sum_{ij} |S_{ij}|^2}. 
\label{totsig}
\ee
In Fig.~\ref{dist} (right) we show the binned distribution 
of the H-asymmetry. As follows from the figure,  
there are regions with different 
signs of asymmetry,  and  size of these regions 
increases with energy. To enhance effect,     
integration over regions with different 
signs of $S$ should be avoided, which requires good enough energy and angle reconstruction. 
Reconstruction of the neutrino energy depends on 
the experimental energy resolution for  muon and 
cascade. Reconstruction of the 
neutrino direction includes both the experimental
and kinematic uncertainties. The latter is related 
to the scattering angle between muon and neutrino. 
These uncertainties can be taken 
into account performing smearing of the distribution
with Gaussian functions characterized by (half) widths
\be
\sigma_E = A E_\nu, ~~~~~\sigma_\theta = B (m_p/E_\nu).   
\ee
Here $A$ and $B$ are the parameters which can be  varied in the intervals 
$A = 0.2 - 0.3$ and $B = 0.3 - 1.5$, $m_p$ is the mass of proton. 
Total significance is given by  formula  
(\ref{totsig}) for smeared $S_{ij}$.  

\begin{figure}
\begin{center}
\includegraphics[width=2.6in]{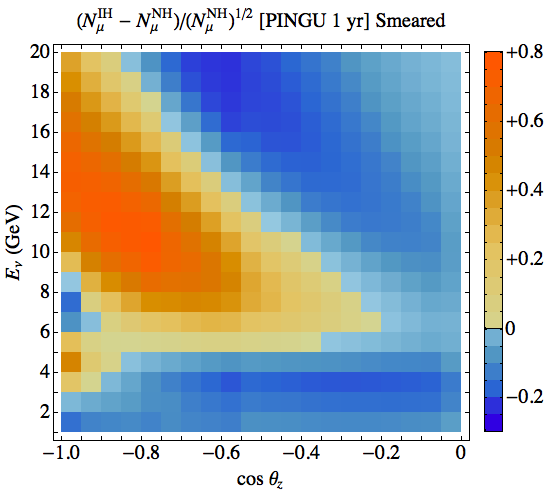}\hskip 4mm
\includegraphics[width=2.6in]{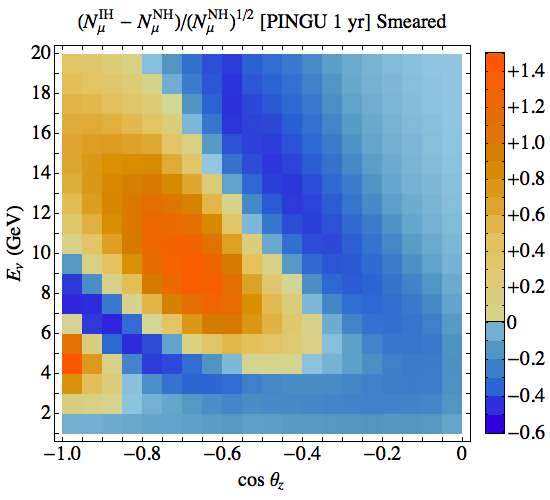}
\end{center}
\caption{Smeared H-asymmetry distribution
of the $\nu_\mu$ events in the $E_\nu - \cos \theta_z$ plane
for two different values of  width of
the angular resolution function: $\sigma_\theta = \sqrt{m_p/ E_\nu}$
({\it left panel}) and $\sigma_\theta = 0.5 \sqrt{m_p/ E_\nu}$
({\it right panel}). We use $\sigma_E = 0.2 E_\nu$ for the width of the energy
reconstruction function (from \cite{ARS}).
}
\label{smearing}
\end{figure}

In Fig.~\ref{smearing} we show the smeared asymmetries 
for two different width $\sigma_\theta$. 
Smearing (i) eliminates fine structures, especially 
at low energies, (ii) shifts the region of  high sensitivity 
to hierarchy to higher energies 
$(8 - 15)$ GeV and larger $|\cos \theta_z|$; 
(iii) reduces significance in the individual bins, and consequently, 
total significance. 
Systematics further reduces significance by factor $\sim 2$ \cite{ARS}.

For parameters  $A = 0.2$ and $B = 1$ (which are rather close to 
the parameters given by PINGU collaboration) one obtains 
that with $f = 5\%$ (uncorrelated systematic errors) the $5\sigma$ 
distinguishability can be achieved in 2 years. The 
distinguishability  
increases up to $7\sigma$ after 5 years of operation. 
These computations correspond to rather large
effective volume. Optimized volume which implements 
an additional cut of events by threshold of 21 DOM hits   
is about 3 time smaller \cite{PINGU} which  further reduces significance 
by $\sim \sqrt{3}$. Still $3\sigma$ distinguishability can be achieved 
after 5 years under conservative assumptions. 
  
Another problem is degeneracy of the hierarchy 
effects with other effects and parameters  involved. 
The strongest degeneracy is related to  
not well known value of $\Delta m_{31}^2$. The smeared 
distribution of quantity  
$A_m \equiv [N_{\mu}^{NH}(\Delta m_{31}^2 + 1\sigma) - 
N_{\mu}^{NH} (\Delta m_{31}^2)]/N_{\mu}^{NH}(\Delta m_{31}^2)$
is very similar to the H-asymmetry distribution. Furthermore, 
maximal significance of $A_m$ in the individual bins 
in about 2 times larger \cite{ARS}. The difference is that 
region of strong $\Delta m_{31}^2$ effect is at 
larger energies $(13 - 18)$ GeV and in vertical direction,  
$\cos \theta_z \approx - 1.0$,  
as compared with (8 - 14) GeV and $\cos \theta_z = - 0.8$ 
for the mass hierarchy. Unfortunately, NOvA and T2K will not be able to 
measure $\Delta m_{31}^2$ with high enough accuracy to avoid 
this degeneracy. Essentially, PINGU will have higher sensitivity 
to $\Delta m_{31}^2$. This means that in analysis 
of the PINGU data one needs to 
consider $\Delta m_{31}^2$  as free fit parameter along 
with the mass hierarchy. 
The $\Delta m_{31}^2-$ dependent asymmetry,  
\be
S_{ij}^2(\Delta m_{31, fit}^2)  = 
[N_{ij}^{NH}(\Delta m_{31, fit}^2) - 
N_{ij}^{NH} (\Delta m_{31, true}^2)]^2/\sigma_{ij}^2, 
\ee
should be minimized with respect to $\Delta m_{31, fit}^2$.  
In the minimum of  $S_{tot}^2(\Delta m_{31}^2)$  
the shift of mass squared difference equals    
$(\Delta m_{31, fit}^2 - \Delta m_{31, true}^2) 
\sim 5 \cdot 10^{-5}$ eV$^2$ which  is about $0.5\sigma$ 
only \cite{ARS}. So, distinguishability is reduced, e.g.,  from 
$6\sigma$ (for fixed $\Delta m_{31}^2$) 
down to $3.8 \sigma$, {\it i.e.} by $37\%$ \cite{ARS}.

\section{Improving sensitivity to mass hierarchy}
\label{impr}

The main challenges for  huge detectors are 

(i) Flavor identification: 
in particular, identification of the $\nu_\mu-$ events in view of  
additional contribution of the   
$\nu_\tau  - \tau  - \mu$  events,  the $\mu - \pi$   misidentification, 
contamination of the $\nu_\mu- $ sample  
by the CC  $\nu_e$ and the neutral current interactions, {\it etc.}.

(ii) Smearing over energies and directions. 
This includes both kinematic smearing, that is,  integration over 
the angle  between the neutrino and muon,  and 
the experimental smearing. 

(iii) Degeneracy of parameters related to uncertainties 
in $\Delta m_{31}^2$, $\theta_{23}$, $\delta$.  
Degeneracy of hierarchy with 
$\delta$ is small~\cite{ARS}: the effect of $\delta$ 
is substantially smaller, and furthermore, it  is mainly at low energies. 

(iv) Systematics. 

\begin{figure}
\includegraphics[width=1.8in]{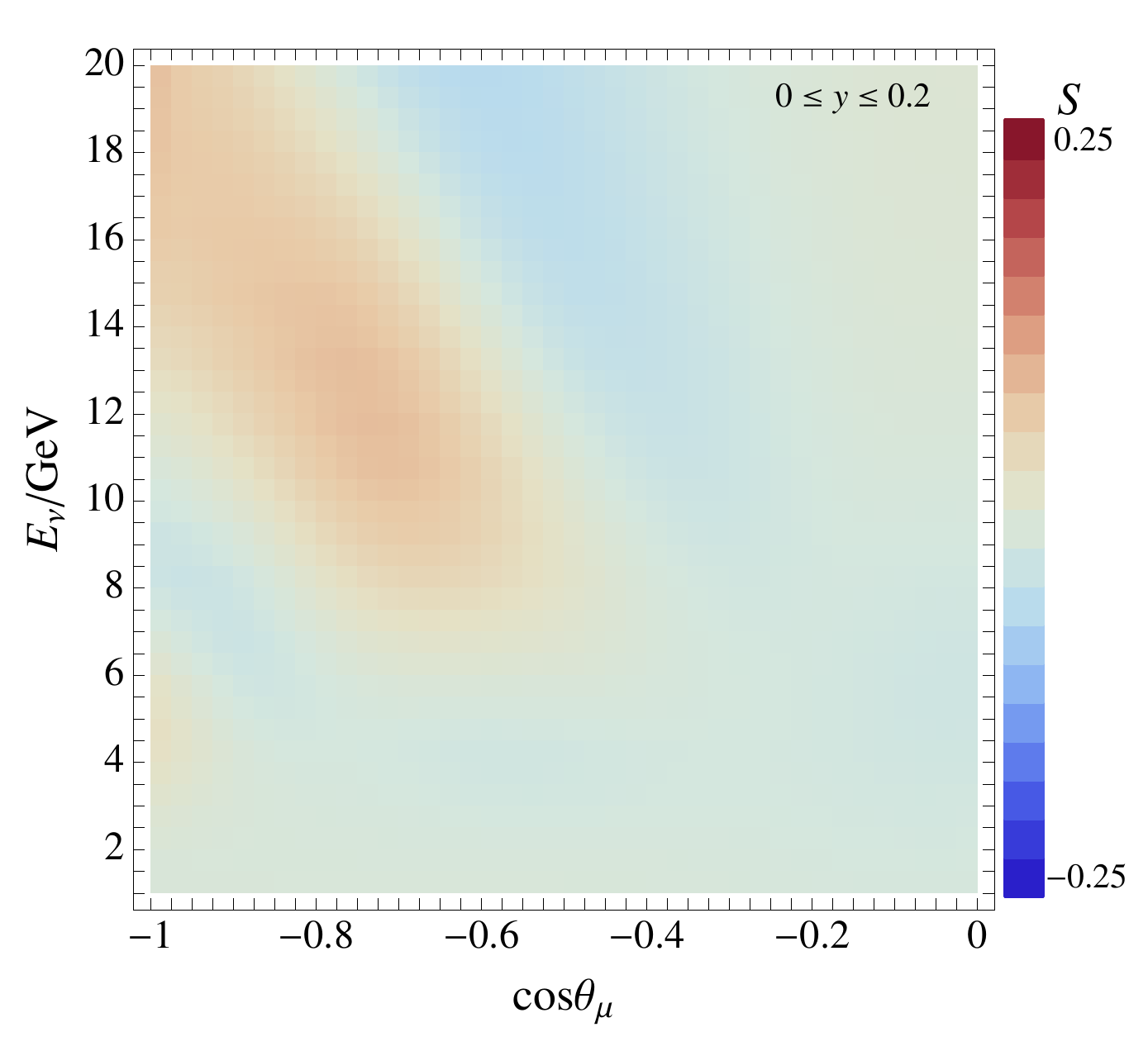} 
\includegraphics[width=1.8in]{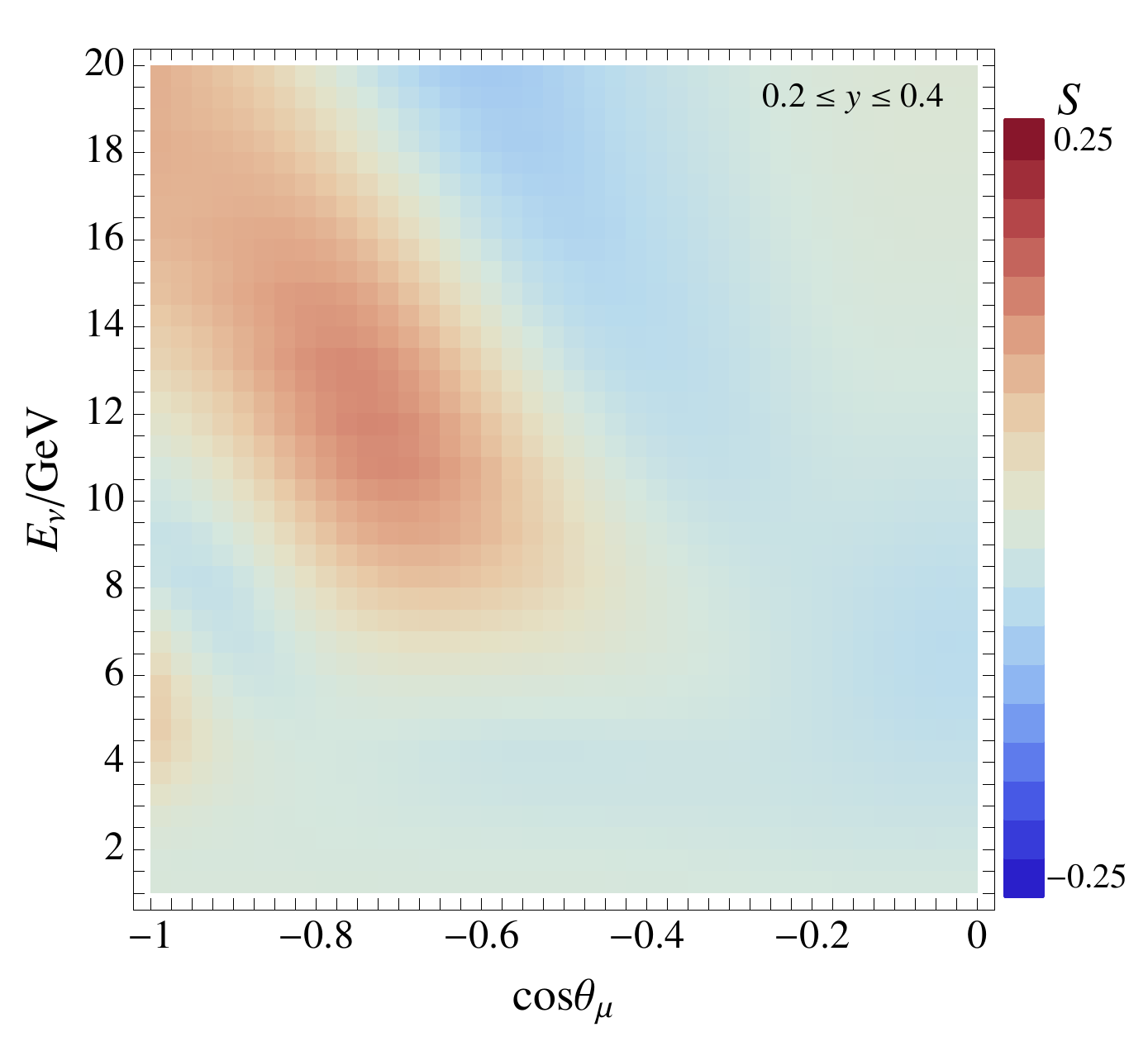}
\includegraphics[width=1.8in]{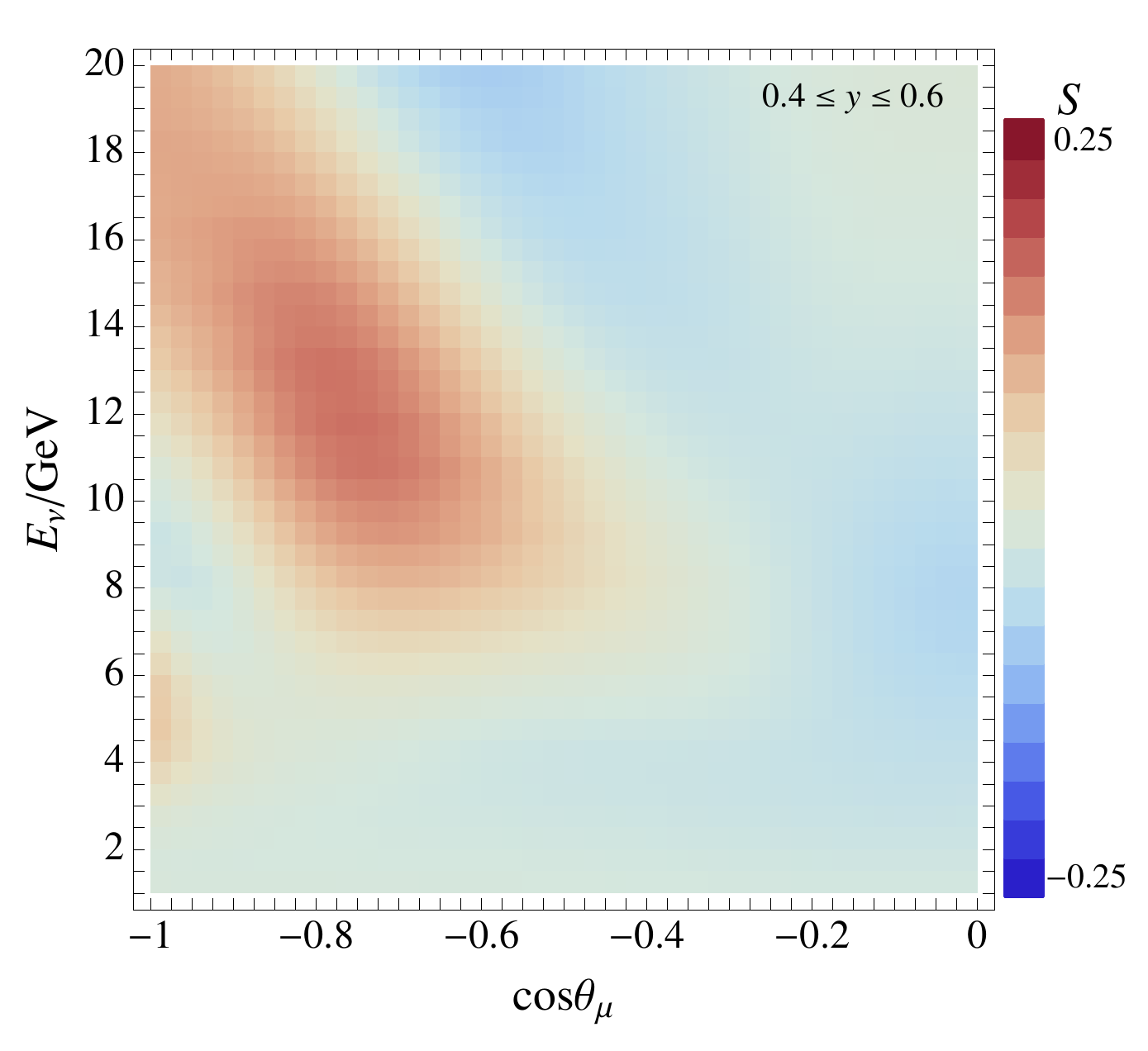}\\ 
\includegraphics[width=1.8in]{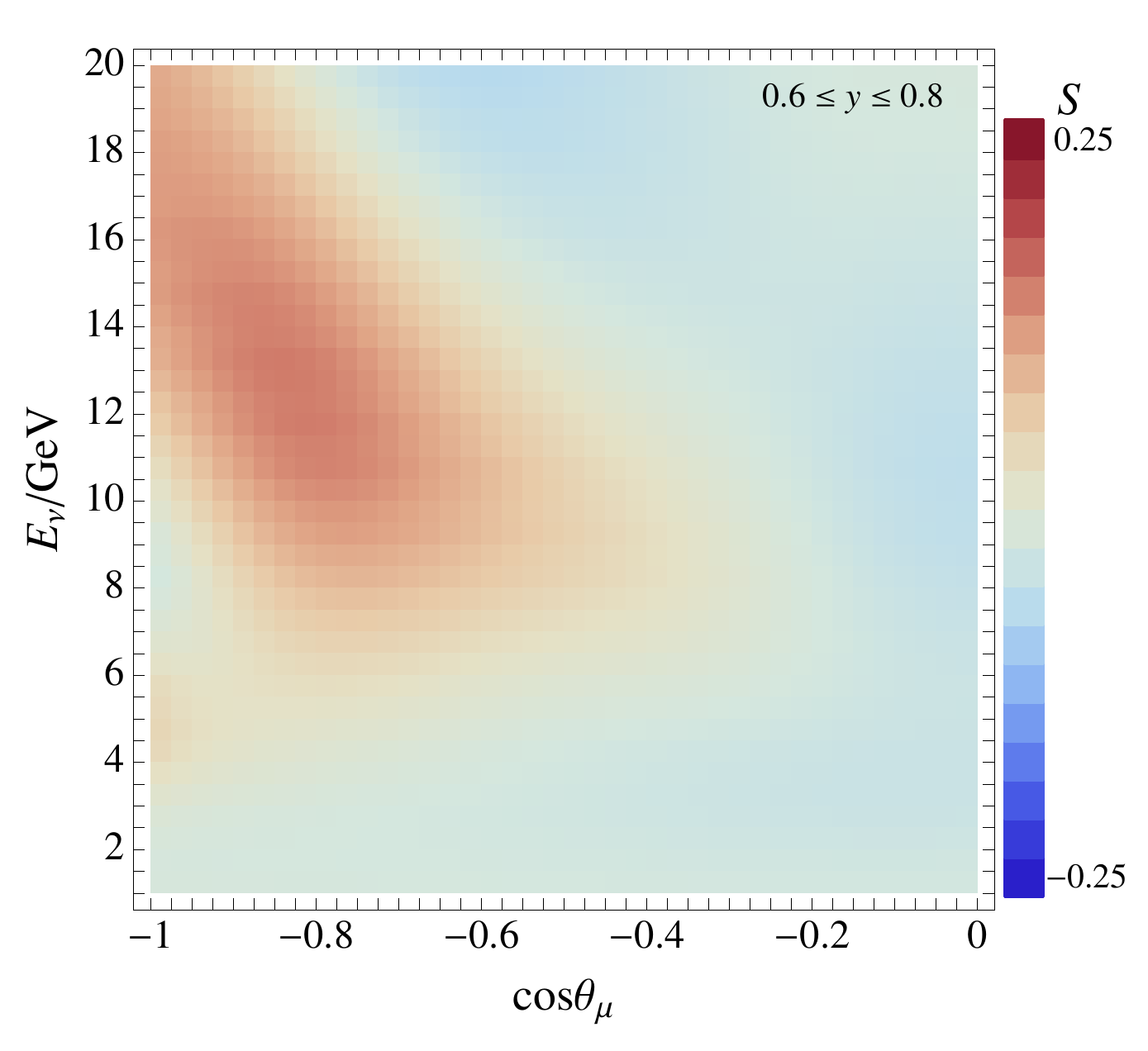}
\includegraphics[width=1.8in]{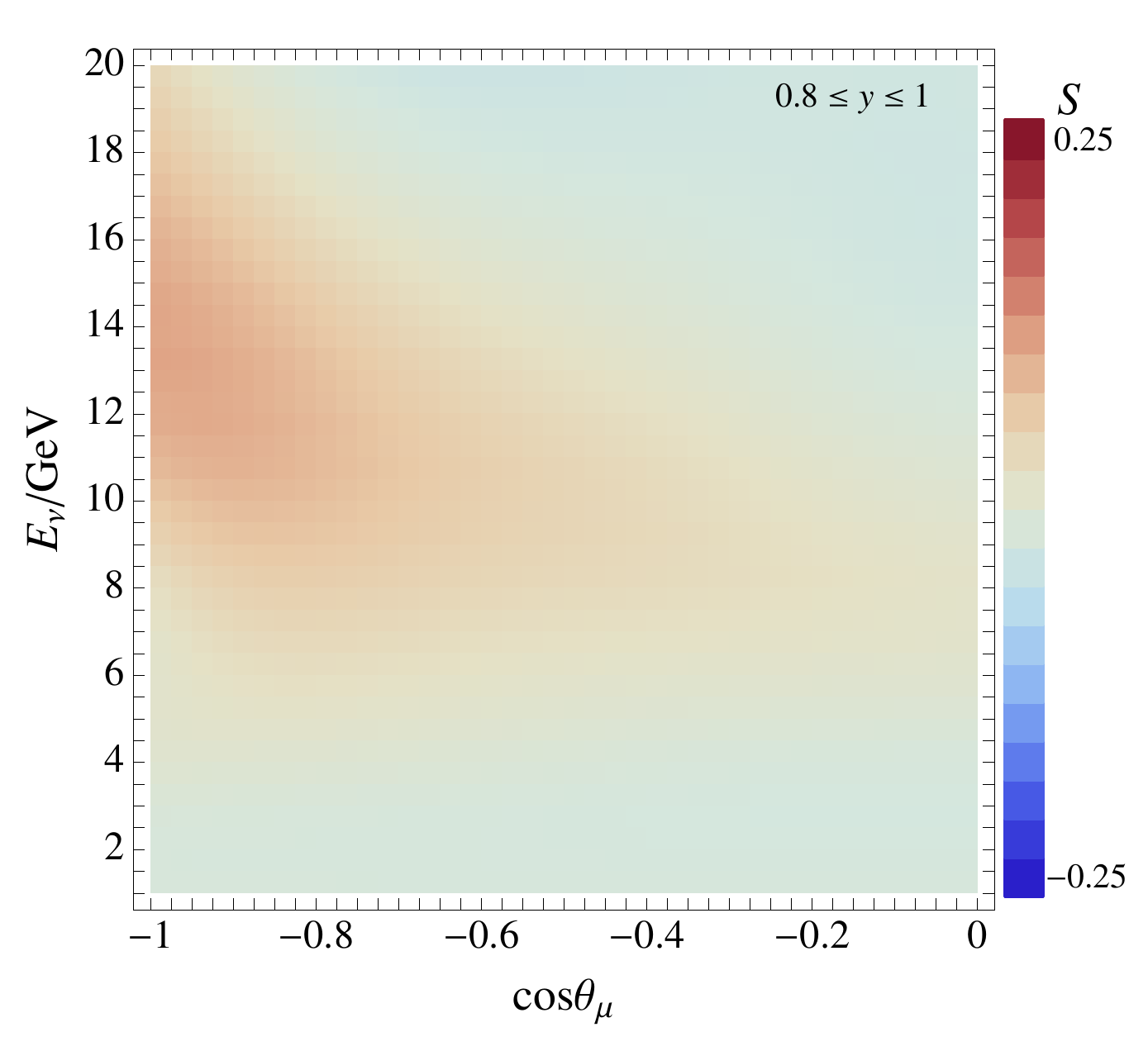}
\includegraphics[width=1.8in]{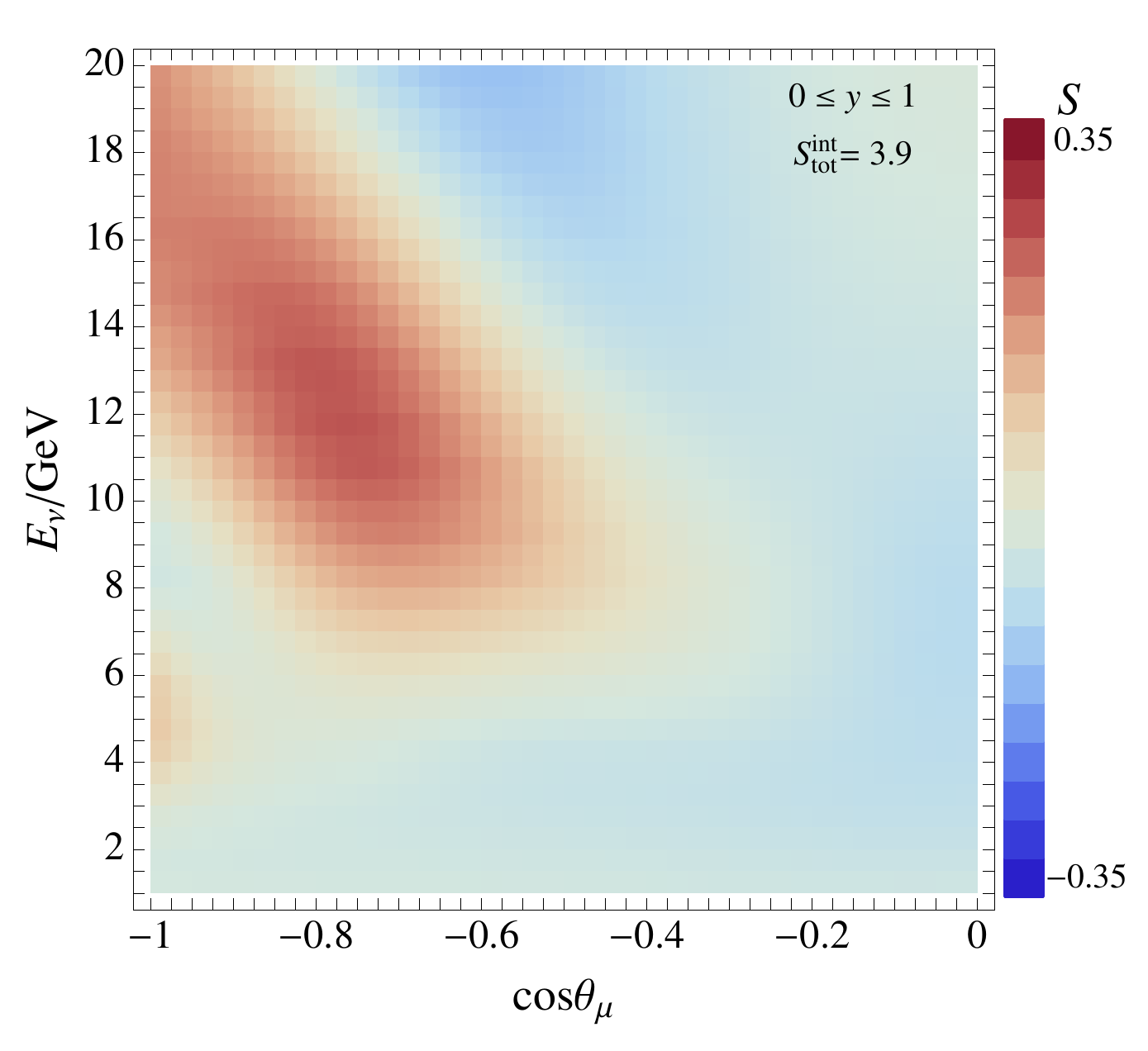}
\caption{
Smeared hierarchy asymmetry distribution
of the $\nu_\mu$ events in the $E_\nu - \cos \theta_z$ plane
for different intervals of $y$. The last panel ({\it bottom, right}) 
shows the $y-$integrated distribution (from \cite{ribordy}).}
\label{ydistr}
\end{figure}

Cancellation of the $\nu$ and $\bar{\nu}$ hierarchy asymmetries reduces the total significance.  
As we have mentioned, if the 
$\nu$ and $\bar{\nu}$ signals are not separated,  they partially cancel each other in 
the total significance: 
\be
S^{tot} \approx |S_\nu| - |\bar{S}_\nu|.  
\ee
In the case  when $\nu$ and $\bar{\nu}$ signals are measured separately, one would have 
\be
S^{tot} \approx \sqrt{S_\nu^2 + z \bar{S}_\nu^2},   
\ee
where $z = 0 (1)$ depends on the 
ratio of numbers of $\nu$ and $\bar{\nu}$  events. 
The ideal separation enhances the significance
by factor 2 - 3. However, 
due to finite accuracy  of separation,   improvement 
of significance is by factor 1.1 - 1.2 only \cite{ribordy}. 

There are different ways to improve sensitivity to hierarchy. 

1) Use the inelasticity ($y-$ distribution)
in addition to $E_\nu$ and $\theta_\mu$, that is,  
the 3 dimensional distributions
of events over $(E_\nu, c_\mu, y)$ or $(E_\mu, E_h,  c_\mu)$, here $c_\mu \equiv \cos \theta_\mu$. 

2) Include other types of events in analysis. 
In particular, the $\nu_e$ events have high sensitivity to the hierarchy \cite{ARS}, \cite{brunner}. 
The problem here is that 
these events have bad angular resolution and they are 
contaminated by the neutral current interactions of all flavors,  
by the $\nu_\tau$ events,   
as well as by the $\nu_\mu$ events with faint muons or muon misidentified with pion. 

3) Direction of cascades.  Already 
$90^{\circ} - 180^{\circ}$ accuracy of the determination would lead 
to substantial improvement of sensitivity to the hierarchy of both  $\nu_e$ and $\nu_\mu$ events.  

Measurements of the inelasticity, {\it i.e.}  
fraction of the total energy transferred to hadrons, 
\be
y = \frac{E_\nu - E_\mu}{E_\nu}, 
\ee
open up a possibility of \cite{ribordy}: 

\begin{itemize} 

\item
Partial separation  (on statistical ground) 
of the neutrino and  antineutrino signals, since  
\be
N_\nu \sim 1, ~~~~ N_\nu \sim (1 - y)^2. 
\ee 
Fitting the measured $y$  distribution for each bin allows one 
to extract fractions of $\nu$ and $\bar{\nu}$.

\item
Better reconstruction of the neutrino direction, 
and therefore reduction of the kinematic smearing: 
Indeed, the angle between neutrino and muon, $\beta$, 
is determined by 
\be
\sin^2 \beta/2 = \frac{x y m_N}{2 E_\mu},  
\ee
where $x$ is the Bjorken variable. 
Selection of events with small $y$ reduces 
possible values of $\beta$.  

\item
Better control over systematics.

\item
Reduction of degeneracy. In fact, 
this gives the largest gain in significance \cite{ribordy}.

\end{itemize}

In Fig.~\ref{ydistr}  we show the H-asymmetries for different intervals of $y$. 
Two improvements, $\nu - \bar{\nu}$ separation and 
reduction of the kinematic smearing anticorrelate:
For small $y$ the angular distribution is good, but  
the $\nu - \bar{\nu}$ separation is bad. On the contrary, 
for large $y$  the $\nu - \bar{\nu}$ separation
improves, but the angular resolution worsens. 
As a result, the largest contribution 
to the significance comes from the intermediate region of $y$. 
Without loss of information and introduction of new systematics 
both effects are taken into account automatically  
if one uses immediately the 3D distribution of events. 
According to Fig. \ref{ydistr} 
with increase of $y$ the region of highest sensitivity shifts 
to larger energies and 
$|\cos \theta_z|$. The change of distribution with $y$ 
allows one to reduce systematics. 

The distribution of events over $E_\nu, c_\mu, y$ 
is given by \cite{ribordy}
\be
n (E_\nu, c_\mu, y) = \frac{1}{\pi} 
\int d c_\mu \rho (E_\nu , c_\nu) g(E_\nu, c_\nu, c_\mu, y), 
\ee
where
\be
\rho (E_\nu , c_\nu) = 2\pi N_A n_{ice} 
V_{eff} T  \Phi_\mu \left[P_{\mu\mu} - r^{-1} P_{e\mu} \right]
\ee
is the density of the neutrino events,  $c_\nu \equiv \cos \theta_z$,  
$r = \Phi_\mu/ \Phi_e$ 
is the ratio of $\nu_\mu$ and $\nu_e$ fluxes, $n_{ice}$ is the density of ice,  
$N_A$ is the Avogadro number, $V_{eff}$ is the effective volume of detector, $T$ 
is the exposure time. Here  
\be
g(E_\nu, c_\nu, c_\mu, y) = \int dx \frac{d^2 \sigma}{dx dy} 
\frac{1}{\sqrt{(s_\mu s_\beta)^2 -(c_\nu - c_\mu c_\beta)^2}}, 
\ee
can be considered as the kinematic smearing function. 
For antineutrinos one needs to change fluxes and the probabilities correspondingly. 
The probabilities (as well as the effective volumes) 
are different for NH and IH.  
Then the  total number of events is given by 
$n = n + \bar{n}$. 
All these distributions should be smeared over the 
experimental resolution functions. 
Total significance of events is given by 
\be
|S_{tot}| = \left\{\int dc_\mu \int dE_\nu \int dy 
\frac{[n^{IH}(E_\nu, c_\mu, y)  - n^{NH}(E_\nu, c_\mu, y)]^2}{n^{NH}(E_\nu, c_\mu, y)}
\right\}^{1/2},  
\label{sss}
\ee
if the $y$-distribution is taken into account. 
Here summation over bins is substituted by integration. 
Without y-distribution (y-integrated) we would have 
\be
|S_{tot}^{int}| = 
\left\{\int dc_\mu \int dE_\nu  
\frac{(\int dy ~ [n^{IH}(E_\nu, c_\mu, y) 
- n^{NH}(E_\nu, c_\mu, y)])^2}{\int dy~ n^{NH}(E_\nu, c_\mu, y)}
\right\}^{1/2} . 
\label{ssstot}
\ee
Notice that in Eq. (\ref{ssstot}) the density of events 
is integrated over $y$   
before computing the significance (under squared).

Results on the total significance (distinguishability) 
which can be obtained during 1 year of exposure 
(for the  effective volume from \cite{ARS}) can be summarized in the following way.
Without experimental  smearing we would have  $S_{tot} = 6.4$.  
Degeneracy reduces this number down to $S_{tot} = 3.6$.  
The use of inelasticity increases it up to $4.8$. 
The  experimental smearing with $\sigma_E = (0.7 E_\nu)^{0.5}$ 
and $\sigma_\psi = 20^{\circ} \sqrt{m_p/E_\mu}$ reduces  these numbers as  
$3.54 \rightarrow 1.90 \rightarrow 2.17$. 


\section{Conclusion}
\label{conc} 

1. Interaction with matter changes the neutrino mixing 
and effective mass splitting 
in a way that depends on the mass hierarchy. 
Consequently, results of  
oscillations and flavor conversion 
are different for the two hierarchies. 

2. Sensitivity to the mass hierarchy appears whenever the 
matter effect on the 1-3 mixing and mass splitting becomes substantial. 
This happens in supernovae in large energy range,  and in the matter of the Earth.  

3. The Earth density profile is a multi-layer medium where the resonance enhancement 
of oscillations as well as the  parametric enhancement 
of oscillations occur. The enhancement  is  realized  in neutrino (antineutrino )  
channels  for normal (inverted) mass hierarchy. 

4. Multi-megaton scale under ice (water) atmospheric  
neutrino detectors with low energy threshold ($2 - 3$ GeV)  may establish mass
hierarchy with $\sim (3 -  10) \sigma$  confidence level in few years.  

5. The main challenges of these experiments are 
flavor identification of events, 
accuracy of measurements of energies and directions, 
systematics, degeneracy of parameters.    
 
6. There are various ways to improve the sensitivity
to the hierarchy.  This includes in particular, 
consideration of the $\nu_e$ events,  reconstruction of the cascade direction, {\it etc.}.    
Inelasticity  measurements will allow to  
increase  significance of the hierarchy 
identification by $20 - 50 \%$ . 
 
7.  Detection of a neutrino burst from relatively close supernovae 
(which may occur any time) may resolve the issue of neutrino 
mass hierarchy. 




\begin{thebibliography}{99}

\bibitem{blennow} 
  M.~Blennow and A.~Y.~ Smirnov,
{\it Neutrino propagation in matter},
  Adv.\ High Energy Phys.\  {\bf 2013} (2013) 972485
  [arXiv:1306.2903 [hep-ph]].

\bibitem{hiermatt}
  J.~Bernabeu, S.~Palomares Ruiz and S.~T.~Petcov,
{\it Atmospheric neutrino oscillations, theta(13) and 
neutrino mass hierarchy,}
  Nucl.\ Phys.\ B {\bf 669} (2003) 255
  [hep-ph/0305152]; 
  D.~Indumathi and M.~V.~N.~Murthy,
{\it A Question of hierarchy: Matter effects with atmospheric 
neutrinos and anti-neutrinos,}
  Phys.\ Rev.\ D {\bf 71} (2005) 013001
  [hep-ph/0407336];
  S.~Palomares-Ruiz and S.~T.~Petcov,
{\it Three-neutrino oscillations of atmospheric neutrinos, 
theta(13), neutrino mass hierarchy and iron magnetized detectors,}
  Nucl.\ Phys.\ B {\bf 712} (2005) 392
  [hep-ph/0406096]; 
  R.~Gandhi, P.~Ghoshal, S.~Goswami, P.~Mehta and S.~U.~Sankar,
{\it Earth matter effects at very long baselines 
and the neutrino mass hierarchy,}
  Phys.\ Rev.\ D {\bf 73} (2006) 053001
  [hep-ph/0411252]; 
  S.~T.~Petcov and T.~Schwetz,
{\it Determining the neutrino mass hierarchy with atmospheric neutrinos,}
  Nucl.\ Phys.\ B {\bf 740} (2006) 1
  [hep-ph/0511277]; 
  R.~Gandhi, P.~Ghoshal, S.~Goswami, P.~Mehta, S.~U.~Sankar and S.~Shalgar,
{\it Mass Hierarchy Determination via future 
Atmospheric Neutrino Detectors,}
  Phys.\ Rev.\ D {\bf 76} (2007) 073012
  [arXiv:0707.1723 [hep-ph]]; 
  V.~Barger, R.~Gandhi, P.~Ghoshal, S.~Goswami, D.~Marfatia, S.~Prakash, S.~K.~Raut and S U.~Sankar,
{\it Neutrino mass hierarchy and octant determination 
with atmospheric neutrinos,}
  Phys.\ Rev.\ Lett.\  {\bf 109} (2012) 091801
  [arXiv:1203.6012 [hep-ph]].  



\bibitem{INO} 
{\it INO, India-Based Neutrino Observatory} 
http://www.ino.tifr.res.in/ino/.  
M.~Blennow and T.~Schwetz,
{\it Identifying the Neutrino mass Ordering with INO and NOvA,}
  JHEP {\bf 1208} (2012) 058
   [Erratum-ibid.\  {\bf 1211} (2012) 098]
  [arXiv:1203.3388 [hep-ph]]; 
  A.~Ghosh, T.~Thakore and S.~Choubey,
{\it Determining the Neutrino Mass Hierarchy with INO, T2K, NOvA 
and Reactor Experiments,}
  JHEP {\bf 1304} (2013) 009
  [arXiv:1212.1305]; 
  A.~Samanta,
{\it Discrimination of mass hierarchy with atmospheric 
neutrinos at a magnetized muon detector,} 
  Phys.\ Rev.\ D {\bf 81} (2010) 037302
  [arXiv:0907.3540 [hep-ph]].  

\bibitem{PINGU} 
  M.~G.~Aartsen {\it et al.}  [IceCube and PINGU Collaborations],
{\it PINGU Sensitivity to the Neutrino Mass Hierarchy,}
  arXiv:1306.5846 [astro-ph.IM], 
  D.~J.~Koskinen,
{\it IceCube-DeepCore-PINGU: Fundamental neutrino
and dark matter physics at the South Pole,}
  Mod.\ Phys.\ Lett.\ A {\bf 26} (2011) 2899.


\bibitem{ORCA} 
P.~ Coyle {\it et al.} {\it Km3Net Contribution to the European Strategy Preparatory 
Group Symposium.} September 2012, Krakow, Poland.  

\bibitem{HK}
  K.~Abe, T.~Abe, H.~Aihara, Y.~Fukuda, Y.~Hayato, K.~Huang, A.~K.~Ichikawa and M.~Ikeda {\it et al.},
{\it Letter of Intent: The Hyper-Kamiokande Experiment --- 
Detector Design and Physics Potential ---,}
  arXiv:1109.3262 [hep-ex].


\bibitem{NOVA} 
  R.~B.~Patterson [NOvA Collaboration],
{\it The NOvA Experiment: Status and Outlook,}
  Nucl.\ Phys.\ Proc.\ Suppl.\  {\bf 235-236} (2013) 151
  [arXiv:1209.0716 [hep-ex]]; 
  O.~Mena, H.~Nunokawa and S.~J.~Parke,
{\it NOvA and T2K: The Race for the neutrino mass hierarchy,}
  Phys.\ Rev.\ D {\bf 75} (2007) 033002
  [hep-ph/0609011].



\bibitem{LBNE} 
  J.~M.~Paley [NOvA and LBNE Collaborations],
{\it The search for CP violation and the determination of the neutrino 
mass hierarchy in NOvA and LBNE,} 
  PoS ICHEP {\bf 2012} (2013) 393.


\bibitem{LBNO} 
  A.~Stahl, C.~Wiebusch, A.~M.~Guler, M.~Kamiscioglu, 
  R.~Sever, A.~U.~Yilmazer, C.~Gunes and D.~Yilmaz {\it et 
  al.},
{\it Expression of Interest for a very long baseline neutrino 
oscillation experiment (LBNO),}
  CERN-SPSC-2012-021.


\bibitem{winter} 
  J.~Tang and W.~Winter,
{\it Requirements for a New Detector at the South Pole Receiving an Accelerator Neutrino Beam,}
  JHEP {\bf 1202} (2012) 028
  [arXiv:1110.5908 [hep-ph]].

\bibitem{parke} 
  H.~Minakata, H.~Nunokawa and S.~J.~Parke,
{\it CP and T violation in neutrino oscillations,}
  AIP Conf.\ Proc.\  {\bf 670} (2003) 132
  [hep-ph/0306221] (see also \cite{ARS}).

\bibitem{solar} 
  S.~Goswami and A.~Y.~ Smirnov,
{\it Solar neutrinos and 1-3 leptonic mixing,}
  Phys.\ Rev.\ D {\bf 72} (2005) 053011
  [hep-ph/0411359].




\bibitem{tomas} 
  M.~Kachelriess and R.~Tomas,
{\it Identifying the neutrino mass hierarchy with supernova neutrinos,} 
  [hep-ph/0412100]; 
  V.~Barger, P.~Huber and D.~Marfatia,
{\it Supernova neutrinos can tell us the neutrino mass 
hierarchy independently of flux models,}
  Phys.\ Lett.\ B {\bf 617} (2005) 167
  [hep-ph/0501184].

\bibitem{fuller} 
  H.~Duan, G.~M.~Fuller, J.~Carlson and Y.~-Q.~Zhong,
{\it Neutrino Mass Hierarchy and Stepwise Spectral Swapping of Supernova Neutrino Flavors,}
  Phys.\ Rev.\ Lett.\  {\bf 99} (2007) 241802
  [arXiv:0707.0290 [astro-ph]];
  B.~Dasgupta, A.~Mirizzi, I.~Tamborra and R.~Tomas,
{\it Neutrino mass hierarchy and three-flavor spectral splits of supernova neutrinos,}
  Phys.\ Rev.\ D {\bf 81} (2010) 093008
  [arXiv:1002.2943 [hep-ph]].


\bibitem{dasgupta} 
  B.~Dasgupta, A.~Dighe, G.~G.~Raffelt and A.~Y.~Smirnov,
  Phys.\ Rev.\ Lett.\  {\bf 103} (2009) 051105
  [arXiv:0904.3542 [hep-ph]].

\bibitem{serpico} 
  P.~D.~Serpico, S.~Chakraborty, T.~Fischer, L.~Hudepohl, H.~-T.~Janka and A.~Mirizzi,
{\it Probing the neutrino mass hierarchy with the rise time of a supernova burst,}
  Phys.\ Rev.\ D {\bf 85} (2012) 085031
  [arXiv:1111.4483 [astro-ph.SR]].

\bibitem{dighe} 
  A.~S.~Dighe and A.~Y.~Smirnov,
{\it Identifying the neutrino mass spectrum from
the neutrino burst from a supernova,}
  Phys.\ Rev.\ D {\bf 62} (2000) 033007
  [hep-ph/9907423].

\bibitem{lunardini} 
  C.~Lunardini and A.~Y.~Smirnov,
{\it Probing the neutrino mass hierarchy
and the 13 mixing with supernovae,}
  JCAP {\bf 0306} (2003) 009
  [hep-ph/0302033].


\bibitem{earth}
  C.~Lunardini and A.~Y.~Smirnov,
{\it Supernova neutrinos: Earth matter effects
and neutrino mass spectrum,}
  Nucl.\ Phys.\ B {\bf 616} (2001) 307
  [hep-ph/0106149]; 
  A.~S.~Dighe, M.~T.~Keil and G.~G.~Raffelt,
{\it Detecting the neutrino mass hierarchy with a supernova at IceCube,}
  JCAP {\bf 0306} (2003) 005
  [hep-ph/0303210].


\bibitem{osc1}
P. Lipari, (1998) unpublished; 
  T.~Ohlsson and H.~Snellman,
{\it Neutrino oscillations with three flavors in matter: 
Applications to neutrinos traversing the Earth,} 
  Phys.\ Lett.\ B {\bf 474} (2000) 153
  [hep-ph/9912295];
T. Kajita {\it Atmospheric neutrinos}, New J. Phys. {\bf 6} (2004) 194. 


\bibitem{maltoni} 
  E.~K.~Akhmedov, M.~Maltoni and A.~Y.~Smirnov,
{\it Neutrino oscillograms of the Earth: Effects of 1-2 mixing
and CP-violation,}
  JHEP {\bf 0806} (2008) 072
  [arXiv:0804.1466 [hep-ph]]; 
  E.~K.~Akhmedov, M.~Maltoni and A.~Y.~Smirnov,
{\it 1-3 leptonic mixing and the neutrino oscillograms of the Earth,}
  JHEP {\bf 0705} (2007) 077
  [hep-ph/0612285].


\bibitem{ARS} 
  E.~K.~Akhmedov, S.~Razzaque and A.~Y.~Smirnov,
{\it Mass hierarchy, 2-3 mixing and CP-phase with
Huge Atmospheric Neutrino Detectors},
  JHEP {\bf 1302} (2013) 082
   [JHEP {\bf 1302} (2013) 082]
   [Erratum-ibid.\  {\bf 1307} (2013) 026]
  [arXiv:1205.7071 [hep-ph]].



\bibitem{param} 
  Q.~Y.~Liu and A.~Y.~Smirnov,
{\it Neutrino mass spectrum with muon-neutrino --->
sterile-neutrino oscillations of atmospheric neutrinos,}
  Nucl.\ Phys.\ B {\bf 524} (1998) 505
  [hep-ph/9712493].
  S.~T.~Petcov,
{\it Diffractive - like (or parametric resonance - like?) enhancement 
of the earth (day - night) effect for solar neutrinos crossing the earth core,}
  Phys.\ Lett.\ B {\bf 434} (1998) 321
  [hep-ph/9805262].
  E.~K.~Akhmedov,
{\it Parametric resonance of neutrino oscillations and passage of solar 
and atmospheric neutrinos through the earth,}
  Nucl.\ Phys.\ B {\bf 538} (1999) 25
  [hep-ph/9805272].
  E.~K.~Akhmedov, A.~Dighe, P.~Lipari and A.~Y.~Smirnov,
{\it Atmospheric neutrinos at Super-Kamiokande
and parametric resonance in neutrino oscillations,}
  Nucl.\ Phys.\ B {\bf 542} (1999) 3
  [hep-ph/9808270].

\bibitem{mena}
  O.~Mena, I.~Mocioiu and S.~Razzaque,
{\it Neutrino mass hierarchy extraction using atmospheric
neutrinos in ice,}
  Phys.\ Rev.\ D {\bf 78} (2008) 093003
  [arXiv:0803.3044 [hep-ph]].

\bibitem{winter-at}
  W.~Winter,
{\it Neutrino mass hierarchy determination with IceCube-PINGU,}
  Phys.\ Rev.\ D {\bf 88} (2013) 013013
  [arXiv:1305.5539 [hep-ph]].
  D.~Franco, C.~Jollet, A.~Kouchner, V.~Kulikovskiy, 
  A.~Meregaglia, S.~Perasso, T.~Pradier and A.~Tonazzo {\it et 
  al.},
{\it Mass hierarchy discrimination with atmospheric neutrinos 
in large volume ice/water Cherenkov detectors,}
  JHEP {\bf 1304} (2013) 008
  [arXiv:1301.4332 [hep-ex]].
  
\bibitem{ribordy} 
  M.~Ribordy and A.~Y.~Smirnov,
{\it Improving the neutrino mass hierarchy
identification with inelasticity measurement in PINGU and ORCA,}
  Phys.\ Rev.\ D {\bf 87} (2013) 113007
  [arXiv:1303.0758 [hep-ph]].

\bibitem{brunner}
  J.~Brunner,
{\it Counting Electrons to Probe the Neutrino Mass Hierarchy,}
  arXiv:1304.6230 [hep-ex].


\end{thebibliography}
\end{document}